\documentclass[
reprint,
pre,
amsmath,amssymb,
aps,
showpacs,
showkeys
]{revtex4-2}

\usepackage{graphicx}
\usepackage{dcolumn}
\usepackage{amsthm}
\usepackage{bm}
\usepackage{array}
\usepackage{booktabs}
\usepackage{multirow}
\usepackage{tikz}
\usetikzlibrary{arrows.meta}
\usepackage{subcaption}

\newtheorem{theorem}{Theorem}

\allowdisplaybreaks

\begin{document}
	
	\preprint{APS/123-QED}
\title{ Exact solution of generalized gauge-invariant Ising chains with multi-spin interactions}
	\author{P.V. Khrapov}
	\email{pvkhrapov@gmail.com}
	\altaffiliation{ORCID: 0000-0002-6269-0727}
	
	\author{S.A. Shchurenkov}
	\email{stepan.shchurenkov201@gmail.com}
	\altaffiliation{ORCID: 0009-0002-4769-3855}
	\affiliation{Independent Researchers, Moscow, Russia}	
	
	\date{\today}
	\begin{abstract}
In this work, exact solutions are obtained for a class of generalized gauge-invariant $n$-chain Ising models ($n=1,2,3,4$) with arbitrary multi-spin interactions that are invariant under the local $\mathbb{Z}_2$ gauge group. On a strip lattice of finite length $L$ and width $n$ with periodic or free boundary conditions, an explicit expression for the partition function is derived using the transfer-matrix method.
		
Two successive transformations are developed: elimination of gauge redundancy and reduction of the original model to an effective $n$-chain Ising model with all possible interactions between neighboring vertical layers. On the basis of the spectral decomposition of the $2^n\times 2^n$ transfer matrix, general formulas are obtained for gauge-invariant correlation functions and Wilson loops of arbitrary width. For $n \le 3$, explicit expressions are derived in terms of eigenvalues and eigenvectors. A detailed analysis of the behavior of the Wilson loop is performed, which allows us to identify regimes exhibiting area-law (confinement-like) and perimeter-law (deconfinement-like) dependence. For specific Hamiltonians, the string tension is computed and the corresponding phase diagrams are constructed. The results generalize and substantially extend the classical works on the gauge-invariant Ising model.
\end{abstract} 
	

	\maketitle

\section{Introduction}
Gauge-invariant lattice models occupy a central place in modern theoretical and mathematical physics as discrete (ultraviolet-regularized) versions of gauge fields, as well as effective descriptions for a wide class of systems in statistical mechanics and condensed-matter physics. In the lattice formulation, gauge symmetry is realized as a local transformation of the degrees of freedom, while the physically observable quantities are gauge-invariant functions of the configuration (for example, products along closed contours). In particular, in works on lattice gauge theories, the confinement criterion has been formulated in terms of the expectation value of Wilson loops: the confinement regime corresponds to a characteristic decay of the average value of the loop with the area (area law), whereas perimeter-law behavior is possible in alternative regimes~\cite{Wilson}.
	
	One of the simplest yet nontrivial examples of a lattice gauge theory is the model with the discrete gauge group $\mathbb{Z}_2$, also known as the gauge-invariant Ising model. Its early formulations and its connection with generalized Ising models and dualities date back to works demonstrating the existence of ``topological'' phase transitions without a local order parameter and with natural gauge-invariant observables~\cite{Wegner}. The development and systematic discussion of the $\mathbb{Z}_2$ case as a lattice gauge theory, including the analysis of the ``temperature-like'' coupling parameter and the presence/type of transition for $d\ge 3$, was carried out, in particular, in a series of works on gauge-invariant fields on lattices, where the gauge-invariant Ising model is specially distinguished~\cite{Balian}.
	
	A general review of the connection between lattice gauge theories and spin systems, as well as the role of the transfer matrix as a key computational tool, is presented in the classical review~\cite{Kogut}. From a fundamental point of view, it is important that local gauge symmetry cannot be spontaneously broken in a formulation without gauge fixing (Elitzur's theorem)~\cite{Elitzur}. This means that the physically meaningful (and universal) characteristics of phases must be expressed through gauge-invariant quantities. Such quantities include, in particular, Wilson loops and correlation functions constructed from gauge-invariant combinations of ``link'' variables. In the context of the gauge-invariant Ising model, a significant contribution to the analysis of the phase structure from mathematically rigorous standpoint was made in Ref.~\cite{Marra_Miracle-Sole}, where convergent expansions are used to obtain results on the phase diagram.
	
	Despite the rich literature on lattice gauge fields, obtaining exact (in the sense of explicit formulas) results for the partition function and gauge-invariant observables in models with nontrivial interactions remains rare. On the other hand, it is well known in statistical mechanics that geometries of the ``strip of finite width and large length'' type often admit exact analysis via finite-dimensional transfer matrices, which makes it possible to compute the free energy and correlation lengths from the leading eigenvalues. For generalized Ising chains and quasi-chains with multi-spin interactions, analogous methods lead to exact results and reveal nontrivial dualities (including self-dualities)~\cite{Turban}. This makes the formulation of gauge-invariant $n$-chain models particularly attractive: on the one hand, the key gauge observables (Wilson loops, gauge-invariant correlators) are preserved; on the other hand, for small $n$ it is possible to obtain explicit spectral formulas.
	
	The present work is devoted to the exact solution of a class of generalized gauge-invariant $n$-chain Ising models with the $\mathbb{Z}_2$ gauge group, which admit a wide set of gauge-invariant multi-spin interactions. A strip lattice of length $L$ and width $n$ with various boundary conditions is considered. For the original system, gauge degrees of freedom are introduced on vertices and ``link'' variables on edges, and the Hamiltonian is given as a sum of gauge-invariant monomials constructed from local link combinations. Then two key transformations are performed: (i) elimination of gauge redundancy, which allows one to express the partition function in terms of independent gauge-invariant link variables, and (ii) reduction to an effective $n$-chain Ising model with all possible multi-spin interactions between neighboring vertical layers. As a result, the problem is reduced to the study of a finite-dimensional transfer matrix of size $2^n\times 2^n$.
	
	The main results are obtained for $n=2,3,4$. Using the transfer-matrix method, an exact expression is obtained for the partition function at finite length $L$ and in the thermodynamic limit $L\to\infty$. In addition, correlation functions of gauge-invariant variables and Wilson loops are studied, which makes it possible to compare various regions of the parameter space, including regimes interpretable as confinement (confinement-like), at the level of explicit formulas for the observables.
	
	The structure of the paper is as follows. In Sec.~\ref{sec:model} the general form of the model on the strip is given, the action of the gauge group is defined, and gauge-invariant link variables are introduced; then two transformations leading to the effective $n$-chain Ising model with multi-spin interactions are formulated. In Sec.~\ref{sec:n1} the one-chain case $n=1$ is considered as an illustration, for which the transfer matrix, correlation function, and expression for the Wilson loop are written out in detail. Sections~\ref{sec:n2}--\ref{sec:n4} are devoted to the cases $n=2,3,4$, respectively, and contain explicit spectral formulas and analysis of the observables (for further comparison of the behavior regimes).
\newpage
\section{General Form of the Model}\label{sec:model}
\subsection{Description of the lattice}
	Consider $n$-chain closed (cyclically closed) lattices of open width and length $L$:
\begin{equation}\label{eq:Lattice Free}
    {V}_{n,L}^{Open}=\{t^m_i; m=0,1,\dots,L-1; i=0,1,\dots,n-1;\,t_i^L\equiv t_i^0\}
\end{equation}
	If we add the condition $t_m^n\equiv t_m^0$, $m=0,1,\dots,n-1$, then we obtain the cyclically closed lattice on the torus:
\begin{equation}\label{eq:Lattice Closed}
    {V}_{n,L}^{Torus}=\{t^m_i; i=0,1,\dots,n-1; \,m=0,1,\dots,L-1; \,t_i^L\equiv t_i^0;\,t_n^m\equiv t_0^m\}
\end{equation}
	In this work, both types of lattices will be considered. If it is inessential whether the lattice has free boundary conditions or is cyclically closed on the torus, we will denote it by $V_{n,L}$.

	We introduce a single ''layer'' of the lattice
\begin{equation}
    V_{n,L}(m) = \{t_i^m, i = 0, \dots, n-1\}, m=0,\dots, L-1.
\end{equation}
	We denote the set of segments (edges) $\{l_{i,j}^m\}$ connecting points $t_i^m$ and $t_j^{m+1}$ as $E_{n,L}(m)$, then
\begin{equation}
    E_{n,L} = \bigcup_{m=0}^{L-1} E_{n,L}(m).
\end{equation}
	Let $U_{n,L}$ be the set of all vertical edges $\{l_i^m\}$ of the layer $V_{n,L}(m)$ connecting $t_i^m$ and $t_{i+1}^m$. Then $U_{n,L} = \bigcup_{m=0}^{L-1} U_{n,L}(m)$ is the set of all vertical intra-layer, transverse edges (Fig.~\ref{fig: scheme}).
\subsection{Gauge space and action of the gauge group}
	Let
\begin{equation}\label{eq:Lambda}
	\Lambda_{n,L}=  V_{n,L} \cup  E_{n,L} \cup  U_{n,L}
\end{equation}

\begin{figure}[h]
\resizebox{0.5\textwidth}{!}{
\begin{tikzpicture}
    \draw (0,0) node[anchor=south west]{$t_0^0$};
    \draw (0,2) node[anchor=north west]{$t_1^0$};
    \draw (2,0) node[anchor=south west]{$t_0^1$};
    \draw (2,2) node[anchor=north west]{$t_1^1$};

    \draw (0,4) node[anchor=south west]{$t_{i}^0$};
    \draw (0,6) node[anchor=north west]{$t_{i+1}^0$};
    \draw (2,4) node[anchor=south west]{$t_{i}^1$};
    \draw (2,6) node[anchor=north west]{$t_{i+1}^1$};

    \draw (0,8) node[anchor=south west]{$t_{j}^0$};
    \draw (0,10) node[anchor=north west]{$t_{j+1}^0$};
    \draw (2,8) node[anchor=south west]{$t_{j}^1$};
    \draw (2,10) node[anchor=north west]{$t_{j+1}^1$};

    \draw (0,12) node[anchor=south west]{$t_{n-1}^0$};
    % \draw (0,14) node[anchor=north west]{$t_{n}^0$};
    \draw (2,12) node[anchor=south west]{$t_{n-1}^1$};
    % \draw (2,14) node[anchor=north west]{$t_{n}^1$};

    \draw (4,0) node[anchor=south west]{$t_0^{m}$};
    \draw (4,2) node[anchor=north west]{$t_1^{m}$};
    \draw (6,0) node[anchor=south west]{$t_0^{m+1}$};
    \draw (6,2) node[anchor=north west]{$t_1^{m+1}$};

    \draw (4.2,4) node[anchor=south west]{$t_{i}^{m}$};
    \draw (4,6) node[anchor=north west]{$t_{i+1}^{m}$};
    \draw (6,4) node[anchor=south west]{$t_{i}^{m+1}$};
    \draw (6,6) node[anchor=north west]{$t_{i+1}^{m+1}$};

    \draw (4,8) node[anchor=south west]{$t_{j}^{m}$};
    \draw (4,10) node[anchor=north west]{$t_{j+1}^{m}$};
    \draw (6,8) node[anchor=south west]{$t_{j}^{m+1}$};
    \draw (6,10) node[anchor=north west]{$t_{j+1}^{m+1}$};

    \draw (4,12) node[anchor=south west]{$t_{n-1}^{m}$};
    % \draw (4,14) node[anchor=north west]{$t_n^{m}$};
    \draw (6,12) node[anchor=south west]{$t_{n-1}^{m+1}$};
    % \draw (6,14) node[anchor=north west]{$t_n^{m+1}$};

    \draw (8,0) node[anchor=south west]{$t_0^{L-1}$};
    \draw (8,2) node[anchor=north west]{$t_1^{L-1}$};
    % \draw (10,0) node[anchor=south west]{$t_0^{L}$};
    % \draw (10,2) node[anchor=north west]{$t_1^{L}$};

    \draw (8,4) node[anchor=south west]{$t_{i}^{L-1}$};
    \draw (8,6) node[anchor=north west]{$t_{i+1}^{L-1}$};
    % \draw (10,4) node[anchor=south west]{$t_{i}^{L}$};
    % \draw (10,6) node[anchor=north west]{$t_{i+1}^{L}$};

    \draw (8,8) node[anchor=south west]{$t_{j}^{L-1}$};
    \draw (8,10) node[anchor=north west]{$t_{j+1}^{L-1}$};
    % \draw (10,8) node[anchor=south west]{$t_{j}^{L}$};
    % \draw (10,10) node[anchor=north west]{$t_{j+1}^{L}$};

    \draw (8,12) node[anchor=south west]{$t_{n-1}^{L-1}$};
    % \draw (8,14) node[anchor=north west]{$t_n^{L-1}$};
    % \draw (10,12) node[anchor=south west]{$t_{n-1}^{L}$};
    % \draw (10,14) node[anchor=north west]{$t_n^{L}$};

    % \draw (0,0) rectangle (2,2);
    % \draw (0,4) rectangle (2,6);
    % \draw (4,4) rectangle (6,6);
    % \draw (4,0) rectangle (6,2);

    \draw (0,0) -- (0,2);
    \draw (2,0) -- (2,2);
    \draw (4,0) -- (4,2);
    \draw (6,0) -- (6,2);
    \draw (8,0) -- (8,2);
    % \draw (10,0) -- (10,2);

    \draw (0,4) -- (0,6);
    \draw (2,4) -- (2,6);
    \draw (4,4) -- (4,6);
    \draw (6,4) -- (6,6);
    \draw (8,4) -- (8,6);
    % \draw (10,4) -- (10,6);

    \draw (0,8) -- (0,10);
    \draw (2,8) -- (2,10);
    \draw (4,8) -- (4,10);
    \draw (6,8) -- (6,10);
    \draw (8,8) -- (8,10);
    % \draw (10,8) -- (10,10);

    % \draw (0,12) -- (0,14);
    % \draw (2,12) -- (2,14);
    % \draw (4,12) -- (4,14);
    % \draw (6,12) -- (6,14);
    % \draw (8,12) -- (8,14);
    % \draw (10,12) -- (10,14);

    \draw [dotted] (0,2) -- (0,4);
    \draw [dotted] (2,2) -- (4,2);
    \draw [dotted] (2,2) -- (2,4);
    % \draw [dotted] (2,2) -- (4,4);
    \draw [dotted] (2,0) -- (4,0);

    \draw [dotted] (4,2) -- (4,4);
    \draw [dotted] (6,2) -- (8,2);
    \draw [dotted] (6,2) -- (6,4);
    % \draw [dotted] (6,2) -- (8,4);
    \draw [dotted] (6,0) -- (8,0);

    \draw [dotted] (8,2) -- (8,4);
    % \draw [dotted] (10,2) -- (10,4);

    \draw [dotted] (0,6) -- (0,8);
    \draw [dotted] (2,6) -- (4,6);
    \draw [dotted] (2,6) -- (2,8);
    % \draw [dotted] (2,6) -- (4,8);
    \draw [dotted] (2,4) -- (4,4);

    \draw [dotted] (4,6) -- (4,8);
    \draw [dotted] (6,6) -- (8,6);
    \draw [dotted] (6,6) -- (6,8);
    % \draw [dotted] (6,6) -- (8,8);
    \draw [dotted] (6,4) -- (8,4);

    \draw [dotted] (8,6) -- (8,8);
    % \draw [dotted] (10,6) -- (10,8);

    \draw [dotted] (0,10) -- (0,12);
    \draw [dotted] (2,10) -- (4,10);
    \draw [dotted] (2,10) -- (2,12);
    % \draw [dotted] (2,10) -- (4,12);
    \draw [dotted] (2,8) -- (4,8);

    \draw [dotted] (4,10) -- (4,12);
    \draw [dotted] (6,10) -- (8,10);
    \draw [dotted] (6,10) -- (6,12);
    % \draw [dotted] (6,10) -- (8,12);
    \draw [dotted] (6,8) -- (8,8);

    \draw [dotted] (8,10) -- (8,12);
    % \draw [dotted] (10,10) -- (10,12);

    \draw [dotted] (2,12) -- (4,12);
    % \draw [dotted] (2,14) -- (4,14);
    \draw [dotted] (6,12) -- (8,12);
    % \draw [dotted] (6,14) -- (8,14);

    \draw (4,4) -- (6,8);
    \draw (5.4,7.1) node[anchor=south]{$l_{i,j}^m$};

    \draw (4,8) -- (6,4);
    \draw (4.6,7.1) node[anchor=south]{$l_{j,i}^m$};

    \draw (4,10) -- (6,12);
    \draw (5.6,11.1) node[anchor=north]{$l_{j+1,n-1}^m$};

    \draw (0,2) -- (2,2);
    \draw (1,2) node[anchor=south]{$l_{1,1}^0$};

    \draw (0,6) -- (2,2);
    \draw (1.25,4.05) node[anchor=south]{$l_{i+1,1}^0$};

    \draw (0,6) -- (2,10);
    \draw (1.5,8) node[anchor=north]{$l_{i+1,j+1}^0$};

    \draw (0,12) -- (2,8);
    \draw (1.2,10.3) node[anchor=south]{$l_{n-1,j}^0$};

    \draw (0,9) node[anchor=east]{$l_{j}^0$};

    \draw (8.15,5) node[anchor=east]{$l_{i}^{L-1}$};

    % \draw [line width=0.35mm, red ](2,2) -- (2,4);
    % \draw [line width=0.35mm, red ](2,4) -- (6,4);
    % \draw [line width=0.35mm, red ](6,2) -- (6,4);
    % \draw [line width=0.35mm, red ](2,2) -- (6,2);

    % \draw [line width=0.35mm, red ](4,12) -- (8,12);
    % \draw [line width=0.35mm, red ](8,8) -- (8,12);
    % \draw [line width=0.35mm, red ](8,8) -- (6,8);
    % \draw [line width=0.35mm, red ](6,8) -- (4,12);

    % \draw [dashed] (0,0) -- (2,2);
    % \draw [dashed] (0,2) -- (2,0);
    % \draw [dashed] (0,4) -- (2,6);
    % \draw [dashed] (2,4) -- (0,6);
    % \draw [dashed] (4,0) -- (6,2);
    % \draw [dashed] (6,0) -- (4,2);
    % \draw [dashed] (4,4) -- (6,6);
    % \draw [dashed] (6,4) -- (4,6);

    \fill[black] (2,0) circle(1.5pt);
    \fill[black] (0,0) circle(1.5pt);
    \fill[black] (2,2) circle(1.5pt);
    \fill[black] (0,2) circle(1.5pt);

    \fill[black] (2,8) circle(1.5pt);
    \fill[black] (0,8) circle(1.5pt);
    \fill[black] (2,10) circle(1.5pt);
    \fill[black] (0,10) circle(1.5pt);

    \fill[black] (2,12) circle(1.5pt);
    \fill[black] (0,12) circle(1.5pt);
    % \fill[black] (2,14) circle(1.5pt);
    % \fill[black] (0,14) circle(1.5pt);

    \fill[black] (6,8) circle(1.5pt);
    \fill[black] (4,8) circle(1.5pt);
    \fill[black] (6,10) circle(1.5pt);
    \fill[black] (4,10) circle(1.5pt);

    \fill[black] (6,12) circle(1.5pt);
    \fill[black] (4,12) circle(1.5pt);
    % \fill[black] (6,14) circle(1.5pt);
    % \fill[black] (4,14) circle(1.5pt);

    % \fill[black] (10,0) circle(1.5pt);
    \fill[black] (8,0) circle(1.5pt);
    % \fill[black] (10,2) circle(1.5pt);
    \fill[black] (8,2) circle(1.5pt);

    % \fill[black] (10,4) circle(1.5pt);
    \fill[black] (8,4) circle(1.5pt);
    % \fill[black] (10,6) circle(1.5pt);
    \fill[black] (8,6) circle(1.5pt);

    % \fill[black] (10,8) circle(1.5pt);
    \fill[black] (8,8) circle(1.5pt);
    % \fill[black] (10,10) circle(1.5pt);
    \fill[black] (8,10) circle(1.5pt);

    % \fill[black] (10,12) circle(1.5pt);
    \fill[black] (8,12) circle(1.5pt);
    % \fill[black] (10,14) circle(1.5pt);
    % \fill[black] (8,14) circle(1.5pt);
    
    \fill[black] (4,0) circle(1.5pt);
    \fill[black] (2,4) circle(1.5pt);
    \fill[black] (4,2) circle(1.5pt);
    \fill[black] (0,4) circle(1.5pt);

    \fill[black] (4,4) circle(1.5pt);
    \fill[black] (6,6) circle(1.5pt);
    \fill[black] (4,6) circle(1.5pt);
    \fill[black] (6,4) circle(1.5pt);

    \fill[black] (2,6) circle(1.5pt);
    \fill[black] (6,2) circle(1.5pt);
    \fill[black] (6,0) circle(1.5pt);
    \fill[black] (0,6) circle(1.5pt);
\end{tikzpicture}}
\caption{Interlayer interaction}
\label{fig: scheme}
\end{figure}
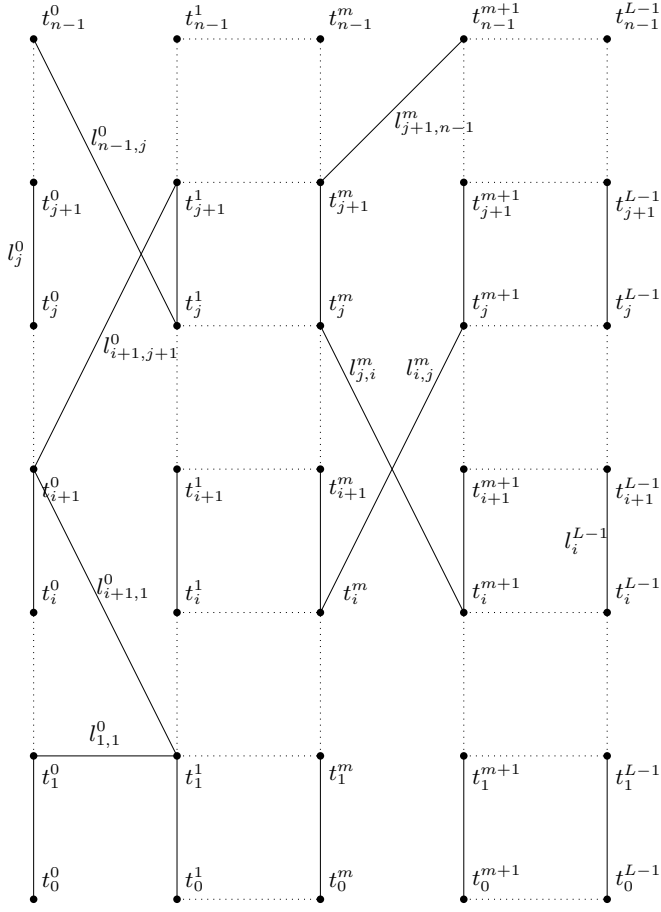

	We assume that on each element of $\Lambda_{n,L}$, i.e., on each vertex of $V_{n,L}$ and on each edge of $E_{n,L}$ and $U_{n,L}$, spins from the space $X=\{+1,-1\}$ are located, respectively: $s_i^m$ on $t_i^m\in V_{n,L}$, $\hat{\gamma}_{i,j}^m$ on $l_{i,j}^m\in E_{n,L}$, and $\hat{\tau}_i^m$ on $l_i^m\in U_{n,L}$ (Fig.~\ref{fig: scheme}).

	The configuration space has the form
\begin{equation}
    \Psi_{n,L} = X^{V_{n,L}} \times X^{U_{n,L}} \times X^{E_{n,L}}.
\end{equation}
	The gauge group $\mathcal{J}=X^{V_{n,L}}$ acts on $\Psi_{n,L}$ in the standard way. For $\kappa=\{\kappa_i^m\}\in\mathcal{J}$,
\begin{equation}
    \begin{aligned}
    & \varkappa = \{\varkappa_{m,i}\}\in \mathcal{J} \\
    &\left(\varkappa \circ s\right)_i^m = \varkappa_i^m \cdot s_i^m \\
    &\left(\varkappa \circ \hat\gamma\right)_{i,j}^m = \varkappa_i^m \cdot \hat\gamma_{i,j}^m \cdot \varkappa_j^{m+1}\\
    &\left(\varkappa \circ \hat\tau\right)_{i}^m = \varkappa_i^m \cdot \hat\tau_{i}^m \cdot \varkappa_{i+1}^{m}
    \end{aligned}
\end{equation}

\subsection{Gauge-invariant ''link'' variables}

	For each edge $l_{i,j}^m\in E_{n,L}$ and $l_i^m\in U_{n,L}$ we define the gauge-invariant ``single-link'' (edge) combination
   \begin{equation}\label{eq:gauge combination}
    \begin{gathered}
   	\tilde{\gamma}_{i,j}^m \left( s, \hat\gamma \right)=s_i^m \hat{\gamma}_{i,j}^m s_j^{m+1}, \\
   	\tilde{\tau}_{i}^m \left( s, \hat\tau \right)=s_i^m \hat{\tau}_{i}^m s_{i+1}^{m}
    \end{gathered}
   \end{equation}
	It follows directly from the definition of the action of $\mathcal{J}$ that
\begin{equation}
    \begin{gathered}
        \tilde\gamma_{i,j}^m \left(\varkappa \circ s,\varkappa \circ \hat\gamma \right) = \gamma_{i,j}^m \left(s,\hat\gamma \right),\\
        \tilde\tau_{i}^m \left(\varkappa \circ s,\varkappa \circ \hat\tau \right) = \tau_{i}^m \left(s,\hat\tau \right),
    \end{gathered}
\end{equation}
	i.e., $\tilde{\gamma}_{i,j}^m$, $\tilde{\tau}_i^m$ are gauge-invariant functions.
   
\subsection{Hamiltonian of the gauge-invariant model under consideration}
	The gauge-invariant functions~\eqref{eq:gauge combination} are uniquely determined by their supports $l_U$ and $l_E$. Let $\omega_U^m\subset U_{n,L}(m)$, $\omega_E^m\subset E_{n,L}(m)$. Then

\begin{equation}
    \begin{gathered}
        \tau_{\omega_U^m} = \prod_{\{\left(m,i\right),\left(m,i+1\right)\}\in\omega_U^m} \tau_i^m ,\\
        \gamma_{\omega_E^m} = \prod_{\{\left(m,i\right),\left(m+1,j\right)\}\in\omega_E^m} \gamma_{i,j}^m,
    \end{gathered}
\end{equation}
	where $\{(m,i),(m+1,j)\}$ is the edge-support of the function $\{\gamma_{i,j}^m\}$ in $E_{n,L}(m)$. Then the $m$-th component of the Hamiltonian~\eqref{eq:gauge combination} can be written as
\begingroup
\footnotesize
\begin{equation}
    \begin{aligned}
    &\hat{\mathcal{H}}_{n,m}(s^m, \hat\tau^m, \hat\gamma^m,s^{m+1},\hat\tau^{m+1}) = \\
    &=\sum_{\tiny(\omega_U^m, \omega_E^m,\omega_U^{m+1}) \subset A} \beta_{\left(\omega_U^m, \omega_E^m, \omega_U^{m+1}\right)} \cdot\\ \cdot&\tau_{\omega_U^m}\left(s^m,\hat\tau^m\right)\gamma_{\omega_E^m}\left(s^m,\hat\gamma^m,s^{m+1}\right)\tau_{\omega_U^{m+1}}\left(s^{m+1},\hat\tau^{m+1}\right),
    \end{aligned}
\end{equation}
\endgroup
	where the summation is over all nonempty subsets $A = U_{n,L}(m)\times E_{n,L}(m)\times U_{n,L}(m+1)$.

	The full Hamiltonian has the form
\begin{equation}
    \mathcal{H}_{n,L}\left(s,\hat\gamma,\hat\tau\right) = \sum_{m=0}^{L-1} \hat{\mathcal{H}}_{n,m}\left(s^m, \hat{\tau}^m, \hat{\gamma}^m, s^{m+1}, \hat{\tau}^{m+1} \right)
\end{equation}
	We assume that the coefficients $\beta(\omega_U^m,\omega_E^m,\omega_U^{m+1})$ are translationally invariant under a cyclic shift along the circle $[0,L-1]$, i.e., they do not depend on the layer $m$. For models on the torus they are also invariant under a cyclic shift in the other direction along $[0,n-1]$, i.e., they do not change under a cyclic shift along the ring $[0,n-1]$.

	Let $\overleftrightarrow{\omega}_E^m = \{l_{j,i}^m\in E_{n,L}(m)\ |\ l_{i,j}^m\in\omega_E^m\}$. Then
\begin{equation}
    \beta_{\left(\omega_U^m, \omega_E^m,\omega_U^{m+1}\right)} = \beta_{\left(\omega_U^{m+1}, \overleftrightarrow\omega_E^m,\omega_U^{m}\right)}
\end{equation}

	We will also call the Hamiltonian symmetric (symmetrical) if $\hat{H}_{n,m}(s^m,\hat{\tau}^m,\hat{\gamma}^m,s^{m+1},\hat{\tau}^{m+1}) = \hat{H}_{n,m}(s^{m+1},\hat{\tau}^{m+1},\overleftrightarrow{\hat{\gamma}}^m,s^m,\hat{\tau}^m)$. It will be shown below that the symmetry of the Hamiltonian leads to symmetry of the transfer matrix, and for simplicity we will consider only symmetric Hamiltonians.

\subsection{Key simplification 1 --- elimination of gauge redundancy} \label{sec: Simplification 1}
	If the Hamiltonian $H_{n,L}(s,\hat{\gamma},\hat{\tau})$ of the model under consideration is a gauge-invariant function, i.e., is in fact expressed as a sum of all possible products of functions of the form~\eqref{eq:gauge combination}, then the change of variables
\begin{equation}
    \{s, \hat\gamma, \hat\tau\} \leftrightarrow \{s, \gamma, \tau\} 
\end{equation}
	is a bijection at fixed $s$. Indeed, from~\eqref{eq:gauge combination}
\begin{equation}\label{eq: One-one relation}
    \begin{gathered}
   	\hat{\gamma}_{i,j}^m =s_i^m \tilde{\gamma}_{i,j}^m s_j^{m+1}, \\
   	\hat{\tau}_{i}^m =s_i^m \tilde{\tau}_{i}^m s_{i+1}^{m}.
    \end{gathered}
\end{equation}
	Consequently, the partition function
    \begin{widetext}
\begin{equation}\label{eq: simplification 1}
    \begin{gathered}
    Z_{n,L} = \sum_{(s, \hat\gamma, \hat\tau)\in\Psi_{n,L}} e^{-\mathcal{H}_{n,L}(s, \hat\gamma, \hat\tau)} = \sum_{s\in X^{V_{n,L}}} \sum_{(\gamma, \tau)\in X^{E_{n,L}}\times X^{U_{n,L}}} e^{-\mathcal{H}_{n,L}(s, \hat\gamma, \hat\tau)} =\\
    =\sum_{s\in X^{V_{n,L}}} \sum_{(\gamma, \tau)\in X^{E_{n,L}}\times X^{U_{n,L}}} e^{-\mathcal{H}^1_{n,L}(\gamma, \tau)} =  2^{|V_{n,L}|} \sum_{(\gamma, \tau)\in X^{E_{n,L}}\times X^{U_{n,L}}} e^{-\mathcal{H}^1_{n,L}(\gamma, \tau)} = 2^{|V_{n,L}|} \cdot Z_{n,L}^1,
    \end{gathered}
\end{equation}
\end{widetext}
	where we pass to the gauge-invariant variables $(\tau,\gamma)\leftarrow(\tilde{\tau},\tilde{\gamma})$, and the Hamiltonian $H_{n,L}^1$ has the form
\begin{widetext}
    \begin{equation}
    \mathcal{H}_{n,L}^1(\gamma,\tau) = \sum_{m=0}^{L-1} \hat{\mathcal{H}}_{n,m}^1\left(\gamma^m\in X^{E_{n,L}(m)}, \tau^m\in X^{U_{n,L}(m)}, \tau^{m+1}\in X^{U_{n,L}(m+1)}\right),
\end{equation}
\end{widetext}
	i.e., it is a sum of terms $\hat{H}_{n,m}^1(\gamma^m,\tau^m,\tau^{m+1})$, each of which depends only on the variables of the $m$-th layer $\gamma_{i,j}^m$, $\tau_i^m$, $\tau_i^{m+1}$.
\begin{equation}
    \hat{\mathcal{H}}_{n,m}^1 = \sum \beta_{\left(\omega_U^m, \omega_E^m, \omega_U^{m+1}\right)} \tau_{\omega_U^m} \gamma_{\omega^m_E} \tau_{\omega_U^{m+1}},
\end{equation}
	where the summation is over all subsets $(\omega_U^m,\omega_E^m,\omega_U^{m+1})\subset U_{n,L}(m)\times E_{n,L}(m)\times U_{n,L}(m+1)$.

	Then the model is equivalent (up to the trivial factor $2^{|V_{n,L}|}$) to a model with a system of independent Ising variables $(\gamma,\tau)$ on the edges, in which a multi-spin interaction is present, determined by the Hamiltonian $H_{n,L}^1$.
\subsection{Key simplification 2 --- reduction to an n-chain Ising model with all possible interactions between spins of neighboring vertical layers}\label{sec: Simplification 2}

	From~\eqref{eq: simplification 1} the reduced partition function $Z_L^1$ has the form
\begin{equation}\label{eq: integration}
    Z_L^1 = \sum_{\tau\in X^{U_{n,L}}}\sum_{\gamma\in X^{E_{n,L}}} e^{-\mathcal{H}_{n,L}^1(\gamma,\tau)}
\end{equation}

	Then the simplified partition function~\eqref{eq: simplification 1} has the form
    \begin{widetext}
\begin{equation}
    \begin{gathered}
        Z_{n,L}^1 = \sum_{\tau\in X^{U_{n,L}}}\sum_{\gamma\in X^{E_{n,L}}} e^{-\mathcal{H}_{n,L}^1(\gamma,\tau)} = \sum_{\tau\in X^{U_{n,L}}} \prod_{m=0}^{L-1} \sum_{\gamma^m\in X^{E_{n,L}(m)}} e^{-\hat{\mathcal{H}}_{n,m}^1(\tau^m,\gamma^m, \tau^{m+1})}= Z_{n,L}^2
    \end{gathered}
\end{equation}
\end{widetext}
	Let
\begin{equation}\label{eq: Hamiltonian 2}
    \hat{\mathcal{H}}^2_{n,m} (\tau^m,\tau^{m+1})= -\ln \left(  \sum_{\gamma^m\in X^{E_{n,L}(m)}} e^{-\hat{\mathcal{H}}_{n,m}^1(\tau^m,\gamma^m, \tau^{m+1})} \right)
\end{equation}

	Then the partition function is represented as the partition function of an Ising model with interaction determined by the described Hamiltonian:
\begin{equation}
    Z_L^1 = \sum_{\tau\in X^{U_{n,L}}} e^{-\sum_{m=0}^{L-1} \hat{\mathcal{H}}_m^2(\tau^m,\tau^{m+1})}
\end{equation} 
	If we now interpret the edges $\tau_i^m$ as sites, we obtain the equivalence of the partition function $Z_L^1$ to the partition function $Z_{n,L}^2$ of the generalized two-dimensional Ising model on an $n\times L$ volume with all possible multi-spin interactions between neighboring layers $U_{n,L}(m)$ and $U_{n,L}(m+1)$.
\subsection{Form of the Hamiltonian of the reduced generalized Ising model with multi-spin interaction} 
	We represent the Hamiltonian $\hat{H}_{n,m}^2(\tau^m,\tau^{m+1})$ in the form
\begin{equation}
    \hat{\mathcal{H}}_{n,m}^2(\tau^m,\tau^{m+1}) = \sum_{\omega \subset \{U_{n,L}(m)\cup U_{n,L}(m+1)\}} K_\omega \tau_\omega,
\end{equation}
	where $\tau_\omega = \prod_{l\in\omega} \tau_l$.
	Then from~\eqref{eq: Hamiltonian 2}
    \begin{widetext}
\begin{equation}
    K_\omega = -\frac{1}{2^{2|U_{n,L}(m)|}}\cdot\sum_{\begin{gathered}\tau^m\in X^{U_{n,L}(m)}, \\ \tau^{m+1}\in X^{U_{n,L}(m+1)}\end{gathered}} \tau_\omega \cdot \ln\left(\sum_{\gamma^m \in X^{E_{n,L}(m)}} e^{-\mathcal{H}_{n,m}^1 (\gamma^m,\tau^m, \tau^{m+1})}\right)
\end{equation}
\end{widetext}
	There are $2^{2|U_{n,L}(m)|}$ subsets in total; we take into account that $|U_{n,L}(m)| = |U_{n,L}(m+1)|$, including $\omega=\emptyset$. Since the Hamiltonian is defined up to a constant, $K_\emptyset$ will not affect the subsequent reasoning.
\hrule
	We also introduce an analog of the free energy, as in Ref.~\cite{Balian},
\begin{equation}
    F_L = -\frac{1}{N_{n,L}} \ln Z, \qquad N_{n,L} = \left|\Lambda_{n,L}\right|
\end{equation}
\hrule
\subsection{Transfer matrix} 	The dimensionality of the problem has been reduced by a factor of $2^{|V_{n,L}|+|E_{n,L}|}$. To investigate the model we apply the transfer-matrix method. We construct the transfer matrix of the $m$-th layer according to the following principle:
\begin{equation}\label{eq:transfer-matrix definition}
    \theta_{i,j} = \sum_{\gamma^m\in X^{E_{n,L}(m)}} e^{-\hat{\mathcal{H}}^1_{m,n} \left(\gamma^m, \tau^m_{i,j},  \tau^{m+1}_{i,j}\right)}
\end{equation}
	where $\tau_i^m$, $\tau_j^{m+1}$ is a specific set of values of the spins of the vertical edges in neighboring layers.

The construction principle is illustrated in \ref{Tab: transfer consctr} for the representative case of the two-chain model.
\begin{table}[!h]
\begin{center}
    \begin{tabular}{|c|c|c|c|c|c|c|}
        \hline
                                &            &  $\tau_0^{m+1}$   &     +      &     +      &      $-$     &      $-$     \\ 
        \hline
                               &            &  $\tau_1^{m+1}$   &     +      &     $-$      &      +     &      $-$    \\ 
       \hline
                    $\tau_0^m$   & $\tau_1^m$   &             &            &            &            &            \\ 
        \hline
         +          & +          &             &     $\theta_{11}$      &     $\theta_{12}$      &      $\theta_{13}$    &      $\theta_{14}$     \\ 
        \hline
         +          & $-$          &             &     $\theta_{21}$      &     $\theta_{22}$      &      $\theta_{23}$     &      $\theta_{24}$     \\
        \hline
         $-$          & +          &             &     $\theta_{31}$      &     $\theta_{32}$      &      $\theta_{33}$     &      $\theta_{34}$     \\
        \hline
         $-$          & $-$          &             &     $\theta_{41}$      &     $\theta_{42}$      &      $\theta_{43}$     &      $\theta_{44}$     \\
        \hline
    \end{tabular}
\end{center}
\caption{Principle of construction of the transfer matrix for the general two-chain model.}
\label{Tab: transfer consctr}
\end{table}\\
We denote ``$+$'' as spin up ($+1$) and ``$-$'' as spin down ($-1$). We alternate the spins for $\tau_i^m$ by $2^{n-i-1}$ spins of one sign, then the opposite, starting from $+1$.
Based on such a configuration of $+1$ and $-1$, we associate with each variable $\tau_i^m$ a spin matrix --- a diagonal matrix with the same configuration of $+1$ and $-1$ on the diagonal. Since all elements of the transfer matrix are positive, its largest eigenvalue in modulus will be positive by the Perron--Frobenius theorem.
\subsection{Wilson loop} 	

In this work we will consider closed contours without self-intersections; the product of the link variables around this contour will be a gauge-invariant function. The Wilson loop is defined as the average of all possible products in this contour.

	For simplicity we will consider contours that are strips of constant width, whose length is oriented along the length $L$ of the lattice (however, the given method allows one to compute the average not only for strip contours but also, for example, for contours of variable width). We denote the horizontal edges entering the Wilson loop contour $W_L^\gamma$.

    In the case when it is necessary to find the average of a product of charges in some contour of length $k$, to take into account the influence of the horizontal edges we will use the matrices obtained in this way. It has the form
\begin{widetext}
\begin{equation}\label{eq:integration with factor}
    \Omega(\tau_i^m,\tau_{i}^{m+1})=\sum\limits_{\{\gamma^m_{i,j}\}=\pm1}\left(\prod\limits_{\gamma_{i,j}^m\in WL_\gamma}\gamma_{i,j}^m\right)\exp\left\{-\mathcal{H}_m^1\left({\tau_i^m,\tau_i^{m+1},\gamma^m}\right)\right\},
\end{equation}
\end{widetext}
	The previously introduced simplification allows one to find explicit expressions for the Wilson loop in an analogous way by changing the measure on the segment from the $i$-th to the $j$-th spins, thereby taking into account all intermediate interactions. The formula takes the form
\begin{equation}\label{eq: Wilson Loop formula}
    WL_{i,j}^\gamma=\frac{Tr\{S_i\Omega_\gamma^{|j-i|}S_j\Theta^{L-|j-i|}\}}{Z_L}
\end{equation}
	where the symmetric matrix $\Omega_\gamma$ is obtained from averaging with allowance for the interaction.
\subsection{Area-law and perimeter-law dependence of the Wilson loop} 	

     By computing the Wilson loop we can find regions in the parameter space of the Hamiltonian in which area-law and perimeter-law dependence of the Wilson loop is observed~\cite{Wilson}.  In models that, unlike the present one, exhibit a phase transition, these two types of scaling indicate the confined and deconfined phases, respectively.

	In the confinement phase, for a strip contour of width $k$, $W_{L k}^{i,j}$ behaves as follows:

\begin{equation}
    WL_{i,j}^k \sim e^{-\sigma S} \Rightarrow WL_{i,j}^k \sim e^{-\sigma k |j-i|}
\end{equation} 
	In the deconfinement phase:
\begin{equation}
    WL_{i,j}^k \sim e^{-bP} \Rightarrow WL_{i,j}^k \sim e^{-b\left( 2k +2|j-i|\right)}
\end{equation}
	The quantity $\sigma$ is called the string tension. One can notice that the exponent of the Wilson loop behaves as $-\sigma k|j-i| + C_1$ or $-2b|j-i| + C_2$, respectively. Both exponents depend linearly on the length of the strip-contour $|j-i|$ for fixed width $k$. We fix $|j-i|$ and consider the difference of the exponents for the Wilson loop over a contour of length $|j-i|$ and $|j-i|+\delta$. We obtain the expressions for area-law and perimeter-law dependence, respectively:
\begin{eqnarray}
    \Delta \ln WL^k = -\sigma k \delta, \label{eq: Area-law}\\
    \Delta \ln WL^k = -2b\delta \label{eq: Perimeter-law}
\end{eqnarray}
	If we consider the ratio of these increments for Wilson loops over contours of widths $k_1$ and $k_2$, we can detect the criterion for area-law and perimeter-law dependence:
\begin{equation}
    \frac{\Delta \ln WL^{k_1}}{\Delta \ln WL^{k_2}} = \begin{cases}
    \displaystyle \frac{k_1}{k_2},  \text{if the dependence is area-law}, \\
    1, \text{if the dependence is perimeter-law}.
    \end{cases}
\end{equation}
	This approach will help to detect regions of the parameter space in which the indicated dependences are observed, as well as to compute the value of the string tension for specific Hamiltonian parameters and contour geometry.

\section{One-Chain Model}\label{sec:n1}
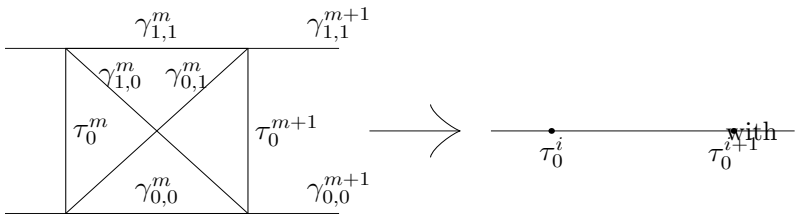
\begin{figure}[h]
\resizebox{300pt}{80pt}{
\begin{tikzpicture}
    \draw (0,0) rectangle (3,3);
    \draw (0,1.5) node[anchor=west]{\Large$\tau_0^m$};
    \draw (1.5,0) node[anchor=south]{\Large$\gamma_{0,0}^{m}$};
    \draw (1.5,3) node[anchor=south]{\Large$\gamma_{1,1}^{m}$};
    \draw (3,1.5) node[anchor=west]{\Large$\tau_0^{m+1}$};
    \draw (4.5,0) node[anchor=south]{\Large$\gamma_{0,0}^{m+1}$};
    \draw (4.5,3) node[anchor=south]{\Large$\gamma_{1,1}^{m+1}$};
    \draw (2,2.1) node[anchor=south]{\Large$\gamma_{0,1}^m$};
    \draw (0.9,2.1) node[anchor=south]{\Large$\gamma_{1,0}^{m}$};

    \draw (-1,0) -- (4.5,0);
    \draw (-1,3) -- (4.5,3);
    \draw (0,0) -- (3,3);
    \draw (0,3)--(3,0);

    \usetikzlibrary {arrows.meta};
    \draw [-{Classical TikZ Rightarrow[length=5mm]}] (5,1.5) -- (6.5,1.5);
    \draw (8,1.5) node[anchor=north]{\Large$\tau_0^i$};
    \draw  (11,1.5) node[anchor=north]{\Large$\tau_0^{i+1}$};
    \fill[black] (11,1.5) circle(1.5pt);
    \fill[black] (8,1.5) circle(1.5pt);
    \draw (7,1.5) -- (12,1.5);
\end{tikzpicture}}
\caption{Transformation of the two-chain gauge-invariant model to the two-chain Ising model.}
\label{fig: one-chain transformation}
\end{figure}

	For a symmetric Hamiltonian the transfer matrix takes the form
\begin{align}\label{eq:transfer-matrix}
    \Theta=
    \begin{pmatrix}
        A  & B \\
        B & C
    \end{pmatrix}
\end{align}
where
\begin{align*}
    &A=\sum\limits_{\{\gamma_{i,j}^m=\pm1\}} e^{-\hat{\mathcal{H}}^1_m(\gamma_{i,j}^m,\tau_0^m=+1,\tau_0^{m+1}=+1)},\\ &C=\sum\limits_{\{\gamma_{i,j}^m=\pm1\}} e^{-\hat{\mathcal{H}}^1_m(\gamma_{i,j}^m,\tau_0^m=-1,\tau_0^{M+1}=-1)},\\
    &B=\sum\limits_{\{\gamma_{i,j}^m=\pm1\}} e^{-\hat{\mathcal{H}}^1_m(\gamma_{i,j}^m,\tau_0^m=+1,\tau_0^{m+1}=-1)}=\\ &=\sum\limits_{\{\gamma_{i,j}^m=\pm1\}} e^{-\hat{\mathcal{H}}^1_m(\gamma_{i,j}^m,\tau_0^m=-1,\tau_0^{m+1}=+1)},
\end{align*}
	with the summation over $\gamma_{i,j}^m$ from the set $\{\gamma_{0,0}^m,\gamma_{0,1}^m,\gamma_{1,0}^m,\gamma_{1,1}^m\}$.
We find its eigenvalues $\lambda_1,\lambda_2$:
\begin{center}
    $\lambda^2-(A+C)\lambda+(AC-B^2)=0$\\
\end{center}
\begin{equation}\label{eq: 1-chain eigenvalues}
    \begin{aligned}
   \displaystyle \lambda_{1,2}=&\frac{A+C\pm\sqrt{(A-C)^2+4B^2}}{2}
    \end{aligned}
\end{equation}
	We have the transfer matrix; now we can write $Z_L$ and the remaining quantities. The partition function of the reduced model is
\begin{align}
    Z_{1,L}^1=Tr \Theta^L=\lambda^L_1+\lambda^L_2
\end{align}
 
\subsection{Correlation function}
	We use the transfer-matrix method to find the average. The formula for the correlation function is
\begin{equation}\label{eq: correlation function formula}
\begin{aligned}
            &G_{ij}=\left<s_0^i\hat\tau_is_{1}^is_{0}^j\hat\tau_js_{1}^j\right>-\left<s_0^i\hat\tau_is_{1}^i\right>\left<s_{0}^j\hat\tau_js_{1}^j\right> =\\ &= \left<\tau_i\tau_j\right>_2-\left<\tau_i\right>_2\left<\tau_j\right>_2 =\\
            &= \frac{Tr\{S\Theta^{|j-i|}S\Theta^{L-|j-i|}\}}{Z_L^1} - \left(\frac{Tr\{\Theta^{i}S\Theta^{L-i}\}}{Z_L^1}\right)^2
\end{aligned}
\end{equation}
where
\begin{equation*}
    S=\begin{pmatrix} 1 & 0 \\ 0 & -1 \end{pmatrix}
\end{equation*}
\begin{equation*}
    \Theta Q=Q\begin{pmatrix} \lambda_1 & 0 \\ 0 & \lambda_2 \end{pmatrix}
\end{equation*}
\begin{equation*}
    Q=\begin{pmatrix} \cos{\varphi} & -\sin{\varphi} \\ \sin{\varphi} & \cos{\varphi} \end{pmatrix}
\end{equation*}

with
\begin{equation*}
    \displaystyle\cos{2\varphi}=\frac{A-C}{\sqrt{(A-C)^2+4B^2}}
\end{equation*}
\begin{equation*}
    \displaystyle\sin{2\varphi}=\frac{2B}{\sqrt{(A-C)^2+4B^2}}
\end{equation*}
	For $\lambda_{1,2}$ found from~\eqref{eq: 1-chain eigenvalues},
\begin{equation*}
    \left<\tau_i\tau_j\right>_2=\cos^2{2\varphi}+\frac{\lambda_2^{|j-i|}\lambda_1^{L-|j-i|}+\lambda_1^{|j-i|}\lambda_2^{L-|j-i|}}{Z_L}\sin^2{2\varphi}    
\end{equation*}
\begin{equation*}
    \displaystyle\left<\tau_i\right>_2=\frac{Tr\{\Theta^{i}S\Theta^{L-i}\}}{Z_L}=\frac{(\lambda_1^L-\lambda_2^L)\cos{2\varphi}}{\lambda_1^L+\lambda_2^L}
\end{equation*}
	Then the correlation function~\eqref{eq: correlation function formula} is found with the help of the expression described in the theorem below.
\begin{widetext}
\begin{theorem}{Correlation function of the one-chain model}
    \begin{equation}\label{eq:1-chain correlation function}
        G_{ij}=\left<\tau_i\tau_j\right>_2-\left<\tau_i\right>_2\left<\tau_j\right>_2=\cos^2{2\varphi}\frac{4\lambda_1^L\lambda_2^L}{(\lambda_1^L+\lambda_2^L)^2}+\frac{\lambda_2^{|j-i|}\lambda_1^{L-|j-i|}+\lambda_1^{|j-i|}\lambda_2^{L-|j-i|}}{\lambda_1^L+\lambda_2^L}\sin^2{2\varphi}
    \end{equation}
		In the limit $L\to\infty$,
    \begin{equation}
        G_{ij}\to\left(\frac{\lambda_2}{\lambda_1}\right)^{|j-i|}\cdot\sin^2{2\varphi}=\sin^2{2\varphi}\cdot e^{ |j-i|\ln{\frac{\lambda_2}{\lambda_1}}}=\sin^2{2\varphi}\cdot e^{-\frac{|j-i|}{\xi}}
    \end{equation}
		where $\xi$ is the correlation length and $m=1/\xi$ is the mass gap.
\end{theorem}
\end{widetext}
\subsection{Wilson loop}
	In the one-chain model it is possible to search only for the average in a strip of width 1:
\begin{widetext}
\begin{equation}
    \begin{aligned}
        &A'=\sum\limits_{\gamma_{0,0}^m=\pm1, \gamma_{1,1}^m=\pm1}\gamma_{0,0}^m\gamma_{1,1}^me^{-\hat{\mathcal{H}}^2_m(\gamma_{0,0}^m,\gamma_{0,1}^m,\gamma_{1,0}^m,\gamma_{0,1}^m,\tau_0^m=+1,\tau_0^{m+1}=+1)}\\
        &B'=\sum\limits_{\gamma_{0,0}^m=\pm1, \gamma_{1,1}^m=\pm1}\gamma_{0,0}^m\gamma_{1,1}^me^{-\hat{\mathcal{H}}^2_m(\gamma_{0,0}^m,\gamma_{0,1}^m,\gamma_{1,0}^m,\gamma_{0,1}^m,\tau_0^m=+1,\tau_0^{m+1}=-1)}=\sum\limits_{\gamma_{0,0}^m=\pm1, \gamma_{1,1}^m=\pm1}\gamma_{0,0}^m\gamma_{1,1}^me^{-\hat{\mathcal{H}}^2_m(\gamma_{0,0}^m,\gamma_{0,1}^m,\gamma_{1,0}^m,\gamma_{0,1}^m,\tau_0^m=-1,\tau_0^{m+1}=+1)}\\
        &C'=\sum\limits_{\gamma_{0,0}^m=\pm1, \gamma_{1,1}^m=\pm1}\gamma_{0,0}^m\gamma_{1,1}^me^{-\hat{\mathcal{H}}^2_m(\gamma_{0,0}^m,\gamma_{0,1}^m,\gamma_{1,0}^m,\gamma_{0,1}^m,\tau_0^m=-1,\tau_0^{m+1}=-1)}
    \end{aligned}
\end{equation}
\end{widetext}
	Then

\begin{align*}
  \Omega=
    \begin{pmatrix}
        A'  & B' \\
        B' & C'
    \end{pmatrix}  
\end{align*}

	We denote the eigenvalues of the matrix $\Omega$ as $\mu_1(=\mu_{\rm max})$, $\mu_2$:
\begin{equation}\label{eq: 1-chain omega eigenvalues}
    \begin{aligned}
            \mu_{1,2}=&\displaystyle\frac{A'+C'\pm\sqrt{(A'-C')^2+4B'^2}}{2}
    \end{aligned}
\end{equation}
	The rotation matrix $R$:
\begin{equation*}
    \Omega R=R \begin{pmatrix} \mu_1 & 0 \\ 0 & \mu_2 \end{pmatrix}
\end{equation*}
\begin{equation*}
    R=\begin{pmatrix} \cos{\psi} & -\sin{\psi} \\ \sin{\psi} & \cos{\psi} \end{pmatrix}    
\end{equation*}
with
\begin{equation*}
    \displaystyle\cos{2\psi}=\frac{A'-C'}{\sqrt{(A'-C')^2+4B'^2}}
\end{equation*}
\begin{equation*}
    \displaystyle\sin{2\psi}=\frac{2B'}{\sqrt{(A'-C')^2+4B'^2}}
\end{equation*}
\begin{equation*}
    \begin{gathered}
        \cos^2(\psi+\varphi)=\cos^2(\varphi)\cos^2(\psi) -2\sin(\varphi)\sin(\psi)\cos(\varphi)\cos(\psi)+ \\+\sin^2(\varphi)\sin^2(\psi)
        =\left(\frac{1+\cos(2\varphi)}{2}\right)\left(\frac{1+\cos(2\psi)}{2}\right)-\\-\frac{1}{2}\sin{2\varphi}\sin{2\psi}+\left(\frac{1-\cos(2\varphi)}{2}\right)\left(\frac{1-\cos(2\psi)}{2}\right)
    \end{gathered}
\end{equation*}
	The following theorem then holds.
\begin{widetext}
\begin{theorem}{Wilson loop of the one-chain model}
    \begin{equation}
        \displaystyle WL_{i,j}=\frac{\left[ \mu_1^{|j-i|}\lambda_1^{L-|j-i|}+\mu_2^{|j-i|}\lambda_2^{L-|j-i|}\right]\cos^2(\psi+\varphi)+\left[ \mu_2^{|j-i|}\lambda_1^{L-|j-i|}+\mu_1^{|j-i|}\lambda_2^{L-|j-i|}\right]\sin^2(\psi+\varphi)}{\lambda_1^L+\lambda_2^L}
    \end{equation}
		In the thermodynamic limit $L\to\infty$,
    \begin{equation}
        \displaystyle WL_{i,j}=\frac{\mu_1^{|j-i|}}{\lambda_1^{|j-i|}}\cos^2(\psi+\varphi)+\frac{\mu_2^{|j-i|}}{\lambda_1^{|j-i|}}\sin^2(\psi+\varphi)
    \end{equation}
		For a long strip $|j-i|\to\infty$,
    \begin{equation}
        \displaystyle WL_{i,j} \to\frac{\mu_1^{|j-i|}}{\lambda_1^{|j-i|}}\cos^2(\psi+\varphi)=\cos^2(\psi+\varphi)\cdot e^{-|j-i|\ln{\frac{\lambda_1}{\mu_1}}}
    \end{equation}
\end{theorem}
\end{widetext}

\newpage
\section{Two-Chain Model}\label{sec:n2}
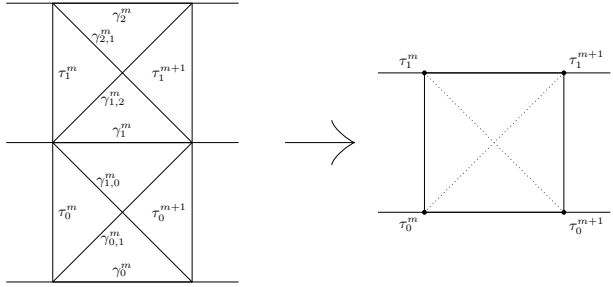
\begin{figure}[!h]
    \centering
\resizebox{0.45\textwidth}{!}{
\begin{tikzpicture}
    \draw (0,0) rectangle (3,3);
    \draw (3,3) rectangle (0,6);
    \draw (0,1.5) node[anchor=west]{$\tau_0^m$};
    \draw (0,4.5) node[anchor=west]{$\tau_1^m$};
    \draw (3,1.5) node[anchor=east]{$\tau_0^{m+1}$};
    \draw (3,4.5) node[anchor=east]{$\tau_1^{m+1}$};
    \draw (1.5,0) node[anchor=south]{$\gamma_0^m$};
    \draw (1.5,3) node[anchor=south]{$\gamma_1^m$};
    \draw (1.5,6) node[anchor=north]{$\gamma_2^m$};
    \draw (1.3,1.2) node[anchor=north]{$\gamma_{0,1}^m$};
    \draw (1.2,1.9) node[anchor=south]{$\gamma_{1,0}^m$};
    \draw (1.3,4.2) node[anchor=north]{$\gamma_{1,2}^m$};
    \draw (1.1,5) node[anchor=south]{$\gamma_{2,1}^m$};

    \draw (-1,0) -- (4,0);
    \draw (-1,3) -- (4,3);
    \draw (-1,6) -- (4,6);
    \draw (0,0) -- (3,3);
    \draw (3,0) -- (0,3);
    \draw (3,3) -- (0,6);
    \draw (0,3) -- (3,6);
    \draw [-{Classical TikZ Rightarrow[length=5mm]}] (5,3) -- (6.5,3);
    \draw (8,1.5) rectangle (11,4.5);
    \draw (8,1.5) node[anchor=north east]{$\tau_0^m$};
    \draw (8,4.5) node[anchor=south east]{$\tau_1^m$};
    \draw (11,1.5) node[anchor=north west]{$\tau_0^{m+1}$};
    \draw (11,4.5) node[anchor=south west]{$\tau_1^{m+1}$};
    \draw (7,1.5) -- (12,1.5);
    \draw (7,4.5) -- (12,4.5);
    \draw [dotted] (8,1.5) -- (11,4.5);
    \draw [dotted] (11,1.5) -- (8,4.5);
    \fill[black] (11,1.5) circle(1.5pt);
    \fill[black] (8,1.5) circle(1.5pt);
    \fill[black] (11,4.5) circle(1.5pt);
    \fill[black] (8,4.5) circle(1.5pt);
\end{tikzpicture}}
\caption{Transformation of the three-chain gauge-invariant model to the two-chain Ising model.}
\end{figure}
\subsection{Partition function}
 For a symmetric Hamiltonian the open-boundary model is equivalent to an Ising model with multi-spin interactions whose transfer matrix has the form
\begin{equation}
    \Theta=\begin{pmatrix} a & b & b & c \\ b & d & f & g \\ b & f & d & g \\ c & g & g & h \end{pmatrix}, \label{eq:2-chain transfer-matrix open}
\end{equation}
	which is obtained by means of the previously defined averaging operation~\eqref{eq: integration}.
	One of the eigenvalues is obvious: $\lambda_4 = d-f$. To find the remaining three eigenvalues it is necessary to solve the cubic equation $\lambda^3 + F\lambda^2 + G\lambda + S = 0$, where
\begin{equation*}
    \begin{aligned}   
        &F=-a - d - f - h\\
        &G=-2 b^2 - c^2 + a d + a f - 2 g^2 + a h + d h + f h\\
        &S=c^2 d + c^2 f - 4 b c g + 2 a g^2 + 2 b^2 h - a d h - a f h
    \end{aligned}
\end{equation*}
	Then the roots are represented in the form~\cite{Cardano}
    \begingroup
    \footnotesize
\begin{equation}\label{eq:2-chain eigenvalues Theta}
    \begin{aligned}
        &\lambda_1=\frac{1}{3}\left(-F+2\sqrt{F^2-3G}\sin\left[\frac{1}{3}\left(\arcsin{\left[\frac{2F^3-9FG+27S}{2(F^2-3G)^{\frac{3}{2}}}\right]}+2\pi\right)\right]\right)\\
        &\lambda_2= \frac{1}{3}\left(-F+2\sqrt{F^2-3G}\sin\left[\frac{1}{3}\left(\arcsin{\left[\frac{2F^3-9FG+27S}{2(F^2-3G)^{\frac{3}{2}}}\right]}+4\pi\right)\right]\right)\\
        &\lambda_3= \frac{1}{3}\left(-F+2\sqrt{F^2-3G}\sin\left[\frac{1}{3}\left(\arcsin{\left[\frac{2F^3-9FG+27S}{2(F^2-3G)^{\frac{3}{2}}}\right]}\right)\right]\right)
    \end{aligned}
\end{equation}
\endgroup
	The largest eigenvalue is $\lambda_1$. Then
\begin{equation}\label{eq: 2-chain partition sum}
    Z_L=\lambda_1^L+\lambda_2^L+\lambda_3^L+\lambda_4^L
\end{equation}
\subsection{Correlation function}
	The diagonalizing matrix for the matrix $\Theta$ is
\renewcommand\arraystretch{1.5}
\begin{equation}\label{eq:2-chain Q open}
     Q=\begin{pmatrix}
            \frac{\alpha_1}{N_1} & \frac{\alpha_2}{N_2} & \frac{\alpha_3}{N_3} & 0 \\ 
            -\frac{\beta_1}{N_1} & -\frac{\beta_2}{N_2} & -\frac{\beta_3}{N_3} & \frac{1}{\sqrt{2}} \\
                -\frac{\beta_1}{N_1} & -\frac{\beta_2}{N_2} & -\frac{\beta_3}{N_3} & -\frac{1}{\sqrt{2}} \\ 
                    \frac{\varepsilon_1}{N_1} & \frac{\varepsilon_2}{N_2} & \frac{\varepsilon_3}{N_3} & 0 
        \end{pmatrix}
\end{equation}
\renewcommand\arraystretch{1}
	where, for nondegenerate eigenvalues,
\begin{equation}\label{eq:2-chain Q expressions}
    \begin{aligned}
    &\alpha_i=\frac{2g^2-(d+f)h+(d+f+h)\lambda_i-\lambda_i^2}{cd+cf-2bg-c\lambda_i},\\
    &\beta_i=\frac{cg-bh+b\lambda_i}{cd+cf-2bg-c\lambda_i},\\
    &\varepsilon_i = 1, \\
    &N_i=\sqrt{\alpha_i^2+2\beta_i^2+\varepsilon_i^2},\,i=1,2,3 
    \end{aligned}
\end{equation}
	In the case $\lambda_2=\lambda_3$ it is necessary to perform orthogonalization, for example
\begin{equation}\label{eq:2-chain Q expressions coincide}
    \begin{aligned}
    &\alpha_2=0, \quad \beta_2 = 1, \quad \varepsilon_2 = -\frac{2b}{c},\\
    &\alpha_3=-\frac{2b}{a-\lambda_3}, \quad \beta_3 = \frac{2b^2}{c^2+2b^2}, \quad \varepsilon_3 = \frac{2bc}{c^2+2b^2},\\
    &N_i=\sqrt{\alpha_i^2+2\beta_i^2+\varepsilon_i^2},\,i=1,2,3 
    \end{aligned}
\end{equation}
\begin{widetext}
\begin{theorem} For the two-chain cyclically closed model in the thermodynamic limit ($L\to\infty$):
		The average value of the spin $\tau_i$:
\begin{equation}\label{eq:2-chain mean charge}
    \left<\tau_i\right>_2=\frac{1}{N_1^2}(\alpha_1^2-\varepsilon_1^2) 
\end{equation}
		The two-spin correlator:
\begin{equation}\label{2-chain correlator}
    \left<\tau_{i,\alpha}\tau_{j,\beta}\right>_2=\frac{Tr\{S_\alpha\Theta^{|j-i|}S_\beta\Theta^{L-|j-i|}\}}{Z_L}
\end{equation}
		If the charges are on the same level ($\alpha=\beta$) at distance $k=|i-j|$ when $L\to\infty$:
\begin{equation}\label{eq:2-chain cor a=b}
    \left<\tau_{i}\tau_{j}\right>_2=2\frac{\beta_1^2}{N_1^2}\left(\frac{\lambda_4}{\lambda_1}\right)^k + \frac{(\alpha_1^2-\varepsilon_1^2)^2}{N_1^4} + \frac{(\alpha_1\alpha_2-\varepsilon_1\varepsilon_2)^2}{N_2^2N_1^2}\left(\frac{\lambda_2}{\lambda_1}\right)^k + \frac{(\alpha_1\alpha_3-\varepsilon_1\varepsilon_3)^2}{N_3^2N_1^2}\left(\frac{\lambda_3}{\lambda_1}\right)^k
\end{equation}
		If the charges are on different levels ($\alpha\neq\beta$) at distance $k=|i-j|$ when $L\to\infty$:
\begin{equation}\label{eq:2-chain cor a!=b}
    \left<\tau_{i}\tau_{j}\right>_2=-2\frac{\beta_1^2}{N_1^2}\left(\frac{\lambda_4}{\lambda_1}\right)^k + \frac{(\alpha_1^2-\varepsilon_1^2)^2}{N_1^4} + \frac{(\alpha_1\alpha_2-\varepsilon_1\varepsilon_2)^2}{N_2^2N_1^2}\left(\frac{\lambda_2}{\lambda_1}\right)^k + \frac{(\alpha_1\alpha_3-\varepsilon_1\varepsilon_3)^2}{N_3^2N_1^2}\left(\frac{\lambda_3}{\lambda_1}\right)^k
\end{equation}
		The correlation functions for one and different levels of the open (not closed on the torus) model in the thermodynamic limit are, respectively,
\begin{equation}
    \begin{gathered}\label{eq:2-chain cor functions}
    \left<\tau_i\tau_j\right>_2-\left<\tau_i\right>_2\left<\tau_j\right>_2 = \frac{(\alpha_1\alpha_2-\varepsilon_1\varepsilon_2)^2}{N_2^2N_1^2}\left(\frac{\lambda_2}{\lambda_1}\right)^k + \frac{(\alpha_1\alpha_3-\varepsilon_1\varepsilon_3)^2}{N_3^2N_1^2}\left(\frac{\lambda_3}{\lambda_1}\right)^k+2\frac{\beta_1^2}{N_1^2}\left(\frac{\lambda_4}{\lambda_1}\right)^k \\
    \left<\tau_i\tau_j\right>_2-\left<\tau_i\right>_2\left<\tau_j\right>_2 = \frac{(\alpha_1\alpha_2-\varepsilon_1\varepsilon_2)^2}{N_2^2N_1^2}\left(\frac{\lambda_2}{\lambda_1}\right)^k + \frac{(\alpha_1\alpha_3-\varepsilon_1\varepsilon_3)^2}{N_3^2N_1^2}\left(\frac{\lambda_3}{\lambda_1}\right)^k-2\frac{\beta_1^2}{N_1^2}\left(\frac{\lambda_4}{\lambda_1}\right)^k
    \end{gathered}
\end{equation}
\end{theorem}
\end{widetext}

\subsection{Wilson loop of width 1}
	Consider the matrix $\Omega_1$, which is the transfer matrix of the model under study obtained after averaging with allowance for the interaction:
\begin{equation}\label{eq:2-chain Omega 1}
\Omega_1=\begin{pmatrix} \omega_{11} & \omega_{12} & \omega_{13} & \omega_{14} \\ \omega_{12} & \omega_{22} & \omega_{23} & \omega_{24} \\ \omega_{13} & \omega_{23} & \omega_{33} & \omega_{34} \\ \omega_{14} & \omega_{24} & \omega_{34} & \omega_{44} \end{pmatrix}
\end{equation}
	where the elements $\omega_{ij}$ can be determined by means of the previously defined averaging operation~\eqref{eq:integration with factor}, with the set over which averaging is performed being the set of horizontal edges $\{\gamma_0^m,\gamma_1^m\}$, or, equivalently, $\{\gamma_1^m,\gamma_2^m\}$. It is symmetric but nothing more; therefore the eigenvalues of $\Omega_1$ are found by solving a characteristic equation of the fourth degree. The rotation matrix will have an arbitrary form; we write the orthogonal matrix $R_1$:
\begin{equation}\label{eq:2-chain R 1}
    R_1=\begin{pmatrix} r_{11} & r_{12} & r_{13} & r_{14} \\ r_{21} & r_{22} & r_{23} & r_{24} \\ r_{31} & r_{32} & r_{33} & r_{34} \\ r_{41} & r_{42} & r_{43} & r_{44} \end{pmatrix}
\end{equation}
	Then one can write the formula for the Wilson loop using the matrix written in the new measure~\eqref{eq:2-chain R 1}, which describes the action between spins at sites $i$ and $j$:
\begin{align}\label{eq:2-chain WL 1 formula}
WL_{i,j}^1=&\frac{Tr\{S_\alpha\Omega^{|j-i|}S_\alpha\Theta^{L-|j-i|}\}}{Z_L}=\nonumber\\
        &=\frac{Tr\{S_\beta\Omega^{|j-i|}S_\beta\Theta^{L-|j-i|}\}}{Z_L}=\nonumber\\
        &=\frac{Tr\{S_\alpha R\boldsymbol{M}^{k} R^{-1}S_\alpha Q{\boldsymbol{\Lambda}}^{L-k} Q^{-1}\}}{Z_L}=\nonumber\\
        &=\frac{Tr\{S_\beta R\boldsymbol{M}^{k} R^{-1}S_\beta Q{\boldsymbol{\Lambda}}^{L-k} Q^{-1}\}}{Z_L},\,\, k=|j-i|
\end{align}
\begin{equation*}
    \begin{aligned}
        \boldsymbol{M}=\begin{pmatrix} \mu_1 & 0 & 0 & 0 \\ 0 & \mu_2 & 0 & 0 \\ 0 & 0 & \mu_3 & 0 \\ 0 & 0 & 0 & \mu_4 \end{pmatrix}\\
        \boldsymbol{\Lambda}=\begin{pmatrix} \lambda_1 & 0 & 0 & 0 \\ 0 & \lambda_2 & 0 & 0 \\ 0 & 0 & \lambda_3 & 0 \\ 0 & 0 & 0 & \lambda_4 \end{pmatrix}
    \end{aligned}
\end{equation*}
\begin{center}
\begin{tikzpicture}
    \draw (0,1.5) -- (0,4.5);
    \draw (3,1.5) -- (3,4.5);
    \draw [dotted] (0.5,1.5) -- (2.5,1.5);
    \draw [dotted] (0.5,4.5) -- (2.5,4.5);
    \draw (0,1.5) node[anchor=north east]{$\tau_{i,\alpha}$};
    \draw (0,4.5) node[anchor=south east]{$\tau_{i,\beta}$};
    \draw (3,1.5) node[anchor=north west]{$\tau_{j,\alpha}$};
    \draw (3,4.5) node[anchor=south west]{$\tau_{j,\beta}$};
    \draw (-1,1.5) -- (0.5,1.5);
    \draw (2.5,1.5) -- (5,1.5);
    \draw (-1,4.5) -- (0.5,4.5);
    \draw (2.5,4.5) -- (5,4.5);
\end{tikzpicture}
\end{center} 
\begin{theorem}
		Explicit expression for the Wilson loop of the two-chain cyclically closed model in the thermodynamic limit ($L\to\infty$), covering a strip of width 1 ($k=|j-i|$):
\begin{equation}\label{eq:2-chain Wilson Loop 1}
    \begin{aligned}
    &WL_{i,j}=\frac{\mu_1^k}{\lambda_1^k}(r_{11}\frac{\alpha_1}{N_1}-r_{21}\frac{\beta_1}{N_1}+r_{31}\frac{\beta_1}{N_1}-r_{41}\frac{\varepsilon_1}{N_1})^2+\\
    &+\frac{\mu_2^k}{\lambda_1^k}(r_{12}\frac{\alpha_1}{N_1}-r_{22}\frac{\beta_1}{N_1}+r_{32}\frac{\beta_1}{N_1}-r_{42}\frac{\varepsilon_1}{N_1})^2+\\
    &+\frac{\mu_3^k}{\lambda_1^k}(r_{13}\frac{\alpha_1}{N_1}-r_{23}\frac{\beta_1}{N_1}+r_{33}\frac{\beta_1}{N_1}-r_{43}\frac{\varepsilon_1}{N_1})^2+\\
    &+\frac{\mu_4^k}{\lambda_1^k}(r_{14}\frac{\alpha_1}{N_1}-r_{24}\frac{\beta_1}{N_1}+r_{34}\frac{\beta_1}{N_1}-r_{44}\frac{\varepsilon_1}{N_1})^2
    \end{aligned}
\end{equation}
\end{theorem}

\subsection{Wilson loop of width 2}
	Consider the matrix $\Omega_2$, which is the transfer matrix of the model under study obtained after averaging with allowance for the interaction~\eqref{eq:integration with factor}, where the multiplication is performed over the set $\{\gamma_0^m,\gamma_2^m\}$:
\begin{equation}\label{eq:2-chain Omega 2}
\Omega_2=\begin{pmatrix} a^* & b^* & b^* & c^* \\ b^* & d^* & f^* & g^* \\ b^* & f^* & d^* & g^* \\ c^* & g^* & g^* & h^* \end{pmatrix}
\end{equation}
	It preserves the symmetries characteristic of the matrix $\Theta$; therefore one can analogously write
\begin{equation}
        \begin{aligned}   
            &F^*=-a^* - d^* - f^* - h^*\\
            &G^*=-2 {b^*}^2 - {c^*}^2 + a^* d^* + a^* f^* - 2 {g^*}^2 + a^* h^* + d^* h^* + f^* h^*\\
            &S^*={c^*}^2 d^* + {c^*}^2 f^* - 4 b^* c^* g^* + 2 a^* {g^*}^2 + \\ &+2 {b^*}^2 h^* - a^* d^* h^* - a^* f^* h^*
        \end{aligned}
\end{equation}
	and the eigenvalues $\mu_{1,2,3,4}$ are found from formulas analogous to~\eqref{eq:2-chain eigenvalues Theta} with the replacement of the coefficients by the starred ones, and $\mu_4=d^*-f^*$.

	The rotation matrix behaves analogously to~\eqref{eq:2-chain Q open}:
\renewcommand\arraystretch{1.5}
\begin{equation}\label{eq:2-chain R 2}
    R_2=\begin{pmatrix} \frac{\gamma_1}{M_1} & \frac{\gamma_2}{M_2} & \frac{\gamma_3}{M_3} & 0 \\ 
        -\frac{\delta_1}{M_1} & -\frac{\delta_2}{M_2} & -\frac{\delta_3}{M_3} & \frac{1}{\sqrt{2}} \\ 
        -\frac{\delta_1}{M_1} & -\frac{\delta_2}{M_2} & -\frac{\delta_3}{M_3} & -\frac{1}{\sqrt{2}} \\ 
        \frac{\zeta_1}{M_1} & \frac{\zeta_2}{M_2} & \frac{\zeta_3}{M_3} & 0
    \end{pmatrix}
\end{equation}
\renewcommand\arraystretch{1}
	with analogous expressions for $\gamma_i$, $\delta_i$, $\zeta_i$, $M_i$ (replacing the coefficients by starred ones).

	Then one can write the formula for the Wilson loop using the new matrix~\eqref{eq:2-chain R 2}, which describes the action between spins at sites $i$ and $j$:
\begin{equation}\label{eq:2-chain WL 2 formula}
    \begin{aligned}
    WL_{i,j}^2=&\frac{Tr\{S_{\alpha\beta}\Omega^{|j-i|}S_{\alpha\beta}\Theta^{L-|j-i|}\}}{Z_L}=\\
            &=\frac{Tr\{S_{\alpha\beta} R\boldsymbol{M}^{k} R^{-1}S_{\alpha\beta} Q{\boldsymbol{\Lambda}}^{L-k} Q^{-1}\}}{Z_L},\,\, k=|j-i|
    \end{aligned}
\end{equation}
where
\begin{equation*}
    \begin{aligned}
        \boldsymbol{M}=\begin{pmatrix} \mu_1 & 0 & 0 & 0 \\ 0 & \mu_2 & 0 & 0 \\ 0 & 0 & \mu_3 & 0 \\ 0 & 0 & 0 & \mu_4 \end{pmatrix}\\
        \boldsymbol{\Lambda}=\begin{pmatrix} \lambda_1 & 0 & 0 & 0 \\ 0 & \lambda_2 & 0 & 0 \\ 0 & 0 & \lambda_3 & 0 \\ 0 & 0 & 0 & \lambda_4 \end{pmatrix}\\
        S_{\alpha\beta}=\begin{pmatrix} 1 & 0 & 0 & 0 \\ 0 & -1 & 0 & 0 \\ 0 & 0 & -1 & 0 \\ 0 & 0 & 0 & 1 \end{pmatrix}
    \end{aligned}
\end{equation*}
\begin{center}
\begin{tikzpicture}
    \draw (0,1.5) -- (0,4.5);
    \draw (3,1.5) -- (3,4.5);
    \draw [dotted] (0.5,1.5) -- (2.5,1.5);
    \draw [dotted] (0.5,4.5) -- (2.5,4.5);
    \draw (0,1.5) node[anchor=north east]{$\tau_{i,\alpha}$};
    \draw (0,4.5) node[anchor=south east]{$\tau_{i,\beta}$};
    \draw (3,1.5) node[anchor=north west]{$\tau_{j,\alpha}$};
    \draw (3,4.5) node[anchor=south west]{$\tau_{j,\beta}$};
    \draw (-1,1.5) -- (0.5,1.5);
    \draw (2.5,1.5) -- (5,1.5);
    \draw (-1,4.5) -- (0.5,4.5);
    \draw (2.5,4.5) -- (5,4.5);
\end{tikzpicture} 
\end{center}
\begin{theorem}
		Explicit expression for the Wilson loop of the two-chain cyclically closed model in the thermodynamic limit ($L\to\infty$), covering a strip of width 2:
\begin{equation}\label{eq:2-chain Wilson Loop 2}
    \begin{aligned}
    WL_{i,j}^2=\frac{\mu_1^k}{\lambda_1^k}&\left(\frac{\gamma_1\alpha_1}{M_1N_1}-2\frac{\delta_1\beta_1}{M_1N_1}+\frac{\varepsilon_1\zeta_1}{M_1N_1}\right)^2 + \\
    &+\frac{\mu_2^k}{\lambda_1^k}\left(\frac{\gamma_2\alpha_1}{M_2N_1}-2\frac{\delta_2\beta_1}{M_2N_1}+\frac{\varepsilon_1\zeta_2}{M_2N_1}\right)^2 + \\
    &+\frac{\mu_3^k}{\lambda_1^k}\left(\frac{\gamma_3\alpha_1}{M_3N_1}-2\frac{\delta_3\beta_1}{M_3N_1}+\frac{\varepsilon_1\zeta_3}{M_3N_1}\right)^2
    \end{aligned}
\end{equation}
		Here one can see that the symmetry of the model gives such a structure of the eigenvector corresponding to $\mu_4$ that the term with this eigenvalue vanishes.
\end{theorem}

\section{Explicit Formulas for Computing the Correlation Function and Wilson Loop of the n-Chain Model}\label{sec: theorems}
	Consider an $n$-chain model of length $L$. In the work we will consider correlation functions of spins of two vertical edges in layers $i$ and $j$; therefore we pass to the generalized Ising model. Let its transfer matrix $\Theta$ have eigenvalues $\lambda_1,\dots,\lambda_{2^n}$ (let $\lambda_1=\lambda_{\rm max}$) and normalized eigenvectors $\vec{v}_1,\dots,\vec{v}_{2^n}$ forming the diagonalizing matrix $Q$. With each function $\tau_\alpha^m$ we associate a diagonal matrix $S_\alpha$. With the product of functions $\tau_{\alpha_1}^m,\dots,\tau_{\alpha_d}^m$ we associate the matrix $S_{[\alpha_1\dots\alpha_d]} = S_{\alpha_1}\cdots S_{\alpha_d}$.

		We introduce the pseudoscalar product
    \begin{equation}
        (\overrightarrow{v},\overrightarrow{\nu})_{S_{\alpha}} = \sum_{i=1}^{2^n}s_i v_i \overline{\nu_i},
    \end{equation}
		where $s_i$ are the diagonal elements of the matrix $S_\alpha$.
\begin{widetext}
\begin{theorem}[Correlation function at finite $L$]
	The two-spin correlator:
    \begin{equation}
        \begin{gathered}
        \left<\tau_\alpha^{m}\tau_\beta^{m+k}\right>_2 = \frac{\operatorname{Tr}\{S_{\alpha} \Theta^{k} S_{\beta} \Theta^{L-k} \}}{Z_L}=\frac{\operatorname{Tr}\{S_{\alpha} Q \boldsymbol{\Lambda}^{k} Q^{-1} S_{\beta} Q \boldsymbol{\Lambda}^{L-k} Q^{-1}\}}{Z_L}=\\
        =\frac{\operatorname{Tr}\{S_{\alpha} (\overrightarrow{v_1}\dots \overrightarrow{v_{2^n}}) \boldsymbol{\Lambda}^{k} (\overrightarrow{v_1}\dots \overrightarrow{v_{2^n}})^{\dagger} S_{\beta} (\overrightarrow{v_1}\dots \overrightarrow{v_{2^n}}) \boldsymbol{\Lambda}^{L-k} (\overrightarrow{v_1}\dots \overrightarrow{v_{2^n}})^{\dagger}\}}{Z_L}=\\
        =\frac{\sum_{i=1}^{2^n}\lambda_i^k \sum_{j=1}^{2^n}\lambda_j^{L-k}\left[\left(\overrightarrow{v_i},\overrightarrow{v_j}\right)_{S_{\alpha}} \left(\overrightarrow{v_j},\overrightarrow{v_i}\right)_{S_{\beta}}\right]}{\sum_{s=1}^{2^n}\lambda_s^L}\\
        \end{gathered}
    \end{equation}
		The average value of one spin:
    \begin{equation}
        \begin{gathered}
        \left<\tau_\alpha\right>_2 = \frac{\operatorname{Tr}\{S_{\alpha} \Theta^{L} \}}{Z_L}=\frac{\operatorname{Tr}\{S_{\alpha} Q \boldsymbol{\Lambda}^{L} Q^{-1} \}}{Z_L}=\frac{\operatorname{Tr}\{S_{\alpha} (\overrightarrow{v_1}\dots \overrightarrow{v_{2^n}}) \boldsymbol{\Lambda}^{L} (\overrightarrow{v_1}\dots \overrightarrow{v_{2^n}})^{\dagger} \}}{Z_L}=\\
        =\frac{\sum_{i=1}^{2^n}\lambda_i^L \left(\overrightarrow{v_i},\overrightarrow{v_i}\right)_{S_{\alpha}} }{\sum_{s=1}^{2^n}\lambda_s^L}\\
        \end{gathered}
    \end{equation}
		The correlation function:
    \begin{equation}
        \begin{gathered}
        G_{\tau_\alpha^{m},\tau_\beta^{m+k}} = \left<\tau_\alpha^{m}\tau_\beta^{m+k}\right>_2 -  \left<\tau_\alpha\right>_2 \left<\tau_\beta\right>_2 = \\
        = \frac{\sum_{i=1}^{2^n} \sum_{j=1}^{2^n}\lambda_i^k\lambda_j^{L-k}\left[\left(\overrightarrow{v_i},\overrightarrow{v_j}\right)_{S_{\alpha}} \left(\overrightarrow{v_j},\overrightarrow{v_i}\right)_{S_{\beta}}\right]}{\sum_{s=1}^{2^n}\lambda_s^L} - \frac{\sum_{i=1}^{2^n}\sum_{j=1}^{2^n}\lambda_i^L \lambda_j^L\left(\overrightarrow{v_i},\overrightarrow{v_i}\right)_{S_{\alpha}} \left(\overrightarrow{v_j},\overrightarrow{v_j}\right)_{S_{\beta}}}{(\sum_{s=1}^{2^n}\lambda_s^L)^2}
        \end{gathered}
    \end{equation}
\end{theorem}
\begin{theorem}[Correlation function at $L\to\infty$]
		The two-spin correlator:
    \begin{equation}
        \begin{gathered}
        \left<\tau_\alpha^{m}\tau_\beta^{m+k}\right>_2 = \frac{\operatorname{Tr}\{S_{\alpha} \Theta^{k} S_{\beta} \Theta^{L-k} \}}{Z_L}=\frac{\operatorname{Tr}\{S_{\alpha} Q \boldsymbol{\Lambda}^{k} Q^{-1} S_{\beta} Q \boldsymbol{\Lambda}^{L-k} Q^{-1}\}}{Z_L}=\\
        =\frac{\operatorname{Tr}\{S_{\alpha} (\overrightarrow{v_1}\dots \overrightarrow{v_{2^n}}) \boldsymbol{\Lambda}^{k} (\overrightarrow{v_1}\dots \overrightarrow{v_{2^n}})^{\dagger} S_{\beta} (\overrightarrow{v_1}\dots \overrightarrow{v_{2^n}}) \boldsymbol{\Lambda}^{L-k} (\overrightarrow{v_1}\dots \overrightarrow{v_{2^n}})^{\dagger}\}}{Z_L}=\\
        =\left(\overrightarrow{v_1},\overrightarrow{v_1}\right)_{S_{\alpha}} \left(\overrightarrow{v_1},\overrightarrow{v_1}\right)_{S_{\beta}} + \frac{\sum_{i=2}^{2^n}\lambda_i^k \left[\left(\overrightarrow{v_i},\overrightarrow{v_1}\right)_{S_{\alpha}} \left(\overrightarrow{v_1},\overrightarrow{v_i}\right)_{S_{\beta}}\right]}{\lambda_1^k}\\
        \end{gathered}
    \end{equation}
		The average value of one spin:
    \begin{equation}
        \begin{gathered}
        \left<\tau_\alpha\right>_2 = \frac{\operatorname{Tr}\{S_{\alpha} \Theta^{L} \}}{Z_L}=\frac{\operatorname{Tr}\{S_{\alpha} Q \boldsymbol{\Lambda}^{L} Q^{-1} \}}{Z_L}=\frac{\operatorname{Tr}\{S_{\alpha} (\overrightarrow{v_1}\dots \overrightarrow{v_{2^n}}) \boldsymbol{\Lambda}^{L} (\overrightarrow{v_1}\dots \overrightarrow{v_{2^n}})^{\dagger} \}}{Z_L}=\\
        =\left(\overrightarrow{v_1},\overrightarrow{v_1}\right)_{S_{\alpha}}\\
        \end{gathered}
    \end{equation}
		The correlation function:
    \begin{equation}
        \begin{gathered}
        G_{\tau_\alpha^{m},\tau_\beta^{m+k}} = \left<\tau_\alpha^{m}\tau_\beta^{m+k}\right>_2 -  \left<\tau_\alpha\right>_2 \left<\tau_\beta\right>_2 = \\
        = \frac{\sum_{i=2}^{2^n}\lambda_i^k \left[\left(\overrightarrow{v_i},\overrightarrow{v_1}\right)_{S_{\alpha}} \left(\overrightarrow{v_1},\overrightarrow{v_i}\right)_{S_{\beta}}\right]}{\lambda_1^k}
        \end{gathered}
    \end{equation}
\end{theorem}
\end{widetext}
    \begin{widetext}
	On the described lattice we consider a closed contour without self-intersections --- a strip of width $d$ stretching from layer $i$ to $j$, $|j-i|=k$. We denote the matrix containing all interlayer interactions entering the contour by $\Omega$; its eigenvalues are $\mu_1,\dots,\mu_{2^n}$ (let $\mu_1=\mu_{\rm max}$) and the corresponding normalized eigenvectors $\vec{r}_1,\dots,\vec{r}_{2^n}$ form the matrix $R$.
\begin{theorem}[Wilson loop at finite $L$]
    \end{theorem}
    \begin{equation}
        \begin{gathered}
        WL_k^d = \frac{\operatorname{Tr}\{S_{\left[\alpha_1\alpha_d\right]} \Omega^{k} S_{\left[\alpha_1\alpha_d\right]}  \Theta^{L-k} \}}{Z_L}=\\
        =\frac{\operatorname{Tr}\{S_{\left[\alpha_1\alpha_d\right]} R \boldsymbol{M}^{k} R^{-1} S_{\left[\alpha_1\alpha_d\right]} Q \boldsymbol{\Lambda}^{L-k} Q^{-1}\}}{Z_L}=\\
        =\frac{\operatorname{Tr}\{S_{\left[\alpha_1\alpha_d\right]} (\overrightarrow{r_1}\dots \overrightarrow{r_{2^n}}) \boldsymbol{M}^{k} (\overrightarrow{r_1}\dots \overrightarrow{r_{2^n}})^{\dagger} S_{\left[\alpha_1\alpha_d\right]} (\overrightarrow{v_1}\dots \overrightarrow{v_{2^n}}) \boldsymbol{\Lambda}^{L-k} (\overrightarrow{v_1}\dots \overrightarrow{v_{2^n}})^{\dagger}\}}{Z_L}=\\
        =\frac{\sum_{i=1}^{2^n}\mu_i^k \sum_{j=1}^{2^n}\lambda_j^{L-k}\left[\left(\overrightarrow{r_i},\overrightarrow{v_j}\right)_{S_{\left[\alpha_1\alpha_d\right]}} \overline{\left(\overrightarrow{r_i},\overrightarrow{v_j}\right)}_{S_{\left[\alpha_1\alpha_d\right]}}\right]}{Z_L}=\\
        =\sum_{i=1}^{2^n}\sum_{j=1}^{2^n}\frac{\mu_i^k\lambda_j^{L-k}}{\sum_{s=1}^{2^n}\lambda_s^L} \left|\left(\overrightarrow{r_i},\overrightarrow{v_j}\right)_{S_{\left[\alpha_1\alpha_d\right]}}\right|^2
        \end{gathered}
    \end{equation}

\begin{theorem}[Wilson loop at $L\to\infty$]
    \end{theorem}
    \begin{equation} \label{eq: Wilson n infty}
        \begin{gathered}
        WL_k^d = \frac{\operatorname{Tr}\{S_{\left[\alpha_1\alpha_d\right]} \Omega^{k} S_{\left[\alpha_1\alpha_d\right]}  \Theta^{L-k} \}}{Z_L}=\\
        =\frac{\operatorname{Tr}\{S_{\left[\alpha_1\alpha_d\right]} R \boldsymbol{M}^{k} R^{-1} S_{\left[\alpha_1\alpha_d\right]} Q \boldsymbol{\Lambda}^{L-k} Q^{-1}\}}{Z_L}=\\
        =\frac{\operatorname{Tr}\{S_{\left[\alpha_1\alpha_d\right]} (\overrightarrow{r_1}\dots \overrightarrow{r_{2^n}}) \boldsymbol{M}^{k} (\overrightarrow{r_1}\dots \overrightarrow{r_{2^n}})^{\dagger} S_{\left[\alpha_1\alpha_d\right]} (\overrightarrow{v_1}\dots \overrightarrow{v_{2^n}}) \boldsymbol{\Lambda}^{L-k} (\overrightarrow{v_1}\dots \overrightarrow{v_{2^n}})^{\dagger}\}}{Z_L}=\\
        =\frac{\lambda_1^{L-k}\sum_{i=1}^{2^n}\mu_i^k \left[\left(\overrightarrow{r_i},\overrightarrow{v_1}\right)_{S_{\left[\alpha_1\alpha_d\right]}} \overline{\left(\overrightarrow{r_i},\overrightarrow{v_1}\right)}_{S_{\left[\alpha_1\alpha_d\right]}}\right]}{\lambda_1^L}=\\
        =\sum_{i=1}^{2^n}\frac{\mu_i^k}{\lambda_1^k} \left|\left(\overrightarrow{r_i},{\overrightarrow{v_1}}\right)_{S_{\left[\alpha_1\alpha_d\right]}}\right|^2
        \end{gathered}
    \end{equation}

\end{widetext}

\section{Three-Chain Model}\label{sec:n3}
\begin{figure}[!h]
    \centering
\resizebox{0.5\textwidth}{!}{
\begin{tikzpicture}
    \draw (0,0) rectangle (3,3);
    \draw (3,3) rectangle (0,6);
    \draw (0,6) rectangle (3,9);

    \draw (0,1.5) node[anchor=west]{$\tau_0^m$};
    \draw (0,4.5) node[anchor=west]{$\tau_1^m$};
    \draw (3,1.5) node[anchor=east]{$\tau_0^{m+1}$};
    \draw (3,4.5) node[anchor=east]{$\tau_1^{m+1}$};
    \draw (0,7.5) node[anchor=west]{$\tau_2^m$};
    \draw (3,7.5) node[anchor=east]{$\tau_2^{m+1}$};
    \draw (1.5,0) node[anchor=south]{$\gamma_0^m$};
    \draw (1.5,3) node[anchor=south]{$\gamma_1^m$};
    \draw (1.5,6) node[anchor=north]{$\gamma_2^m$};
    \draw (1.5,9) node[anchor=north]{$\gamma_3^m$};
    \draw (1.3,1.2) node[anchor=north]{$\gamma_{0,1}^m$};
    \draw (1.2,1.9) node[anchor=south]{$\gamma_{1,0}^m$};
    \draw (1.3,4.2) node[anchor=north]{$\gamma_{1,2}^m$};
    \draw (1.1,5) node[anchor=south]{$\gamma_{2,1}^m$};
    \draw (1.3,7.2) node[anchor=north]{$\gamma_{2,3}^m$};
    \draw (1.1,8) node[anchor=south]{$\gamma_{3,2}^m$};

    \draw (-1,0) -- (1,0);
    \draw (3,0) -- (4,0);
    \draw (-1,3) -- (0,3);
    \draw (3,3) -- (4,3);
    \draw (-1,6) -- (0,6);
    \draw (3,6) -- (4,6);
    \draw (-1,9) -- (0,9);
    \draw (3,9) -- (4,9);
    \draw (0,0) -- (3,3);
    \draw (3,0) -- (0,3);
    \draw (3,3) -- (0,6);
    \draw (0,3) -- (3,6);
    \draw (0,6) -- (3,9);
    \draw (3,6) -- (0,9);
    \draw [-{Classical TikZ Rightarrow[length=5mm]}] (5,4.5) -- (6.5,4.5);

    \draw (8,1.5) rectangle (11,4.5);
    \draw (8,4.5) rectangle (11,7.5);
    \draw (8,1.5) node[anchor=north east]{$\tau_0^m$};
    \draw (8,4.5) node[anchor=south east]{$\tau_1^m$};
    \draw (11,1.5) node[anchor=north west]{$\tau_0^{m+1}$};
    \draw (11,4.5) node[anchor=south west]{$\tau_1^{m+1}$};
    \draw (8,7.5) node[anchor=north east]{$\tau_2^m$};
    \draw (11,7.5) node[anchor=north west]{$\tau_2^{m+1}$};
    \draw (7,1.5) -- (8,1.5);
    \draw (11,1.5) -- (12,1.5);
    \draw (7,4.5) -- (8,4.5);
    \draw (11,4.5) -- (12,4.5);
    \draw (7,7.5) -- (8,7.5);
    \draw (11,7.5) -- (12,7.5);
    \draw [dotted] (8,1.5) -- (11,4.5);
    \draw [dotted] (11,1.5) -- (8,4.5);
    \draw [dotted] (8,4.5) -- (11,7.5);
    \draw [dotted] (11,4.5) -- (8,7.5);
    \fill[black] (11,1.5) circle(1.5pt);
    \fill[black] (8,1.5) circle(1.5pt);
    \fill[black] (11,4.5) circle(1.5pt);
    \fill[black] (8,4.5) circle(1.5pt);
    \fill[black] (11,7.5) circle(1.5pt);
    \fill[black] (8,7.5) circle(1.5pt);
\end{tikzpicture}}
\caption{Transformation of the four-chain gauge-invariant model to the three-chain Ising model.}
\label{fig: three-chain transform}
\end{figure}
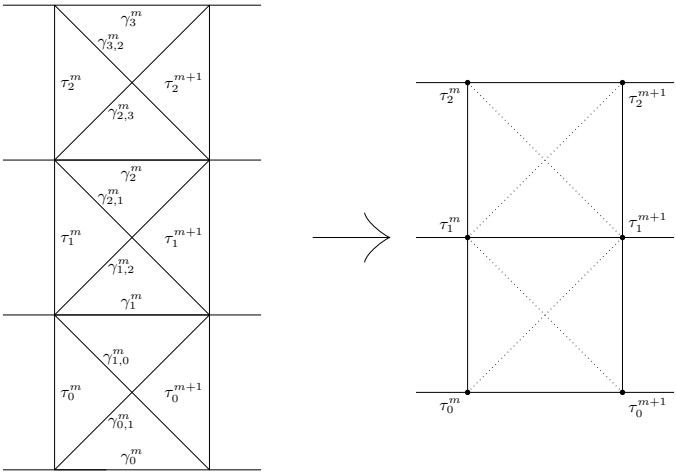
\subsection{Closed model}
\subsubsection{Partition function}
	For a symmetric Hamiltonian the closed model is equivalent to an Ising model with multi-spin interaction determined by the transfer matrix
\begin{equation}
    \Theta=
    \begin{pmatrix}
            a      &     b      &      b     &      c     &     b      &      c     &      c     &      d     \\ 
            b      &     e      &      f     &      g     &     f      &      h     &      i     &      j     \\
            b      &     f      &      e     &      h     &     f      &      i     &      g     &      j     \\
            c      &     g      &      h     &      k     &     i      &      l     &      l     &      m     \\
            b      &     f      &      f     &      i     &     e      &      g     &      h     &      j     \\
            c      &     h      &      i     &      l     &     g      &      k     &      l     &      m     \\
            c      &     i      &      g     &      l     &     h      &      l     &      k     &      m     \\
            d      &     j      &      j     &      m     &     j      &      m     &      m     &      n     
    \end{pmatrix}
\end{equation}
	This model has been considered in Ref.~\cite{Volkov_Khrapov}. First, this shows that the described simplifications really allow one to pass to an Ising model with multi-spin interaction; second, the formulas described in Sec.~\ref{sec: theorems} generalize the formulas given in Ref.~\cite{Volkov_Khrapov} for computing correlation functions.

\subsubsection{Correlation function}

	The form of the correlation function is:

	For spins lying on one chain $i=0,1,2$:
\begin{equation}
    G_{\tau_i^m,\tau_i^{m+k}} = \frac{\sum_{i=2}^{2^n}\lambda_i^k \left[\left(\overrightarrow{v_i},\overrightarrow{v_1}\right)_{S_i} \left(\overrightarrow{v_1},\overrightarrow{v_i}\right)_{S_i}\right]}{\lambda_1^k},
\end{equation}
	For spins lying on neighboring chains, where the lower indices can be subjected to the permutations $(012)$ and $(210)$:
\begin{equation}
    G_{\tau_0^m,\tau_{1}^{m+k}} = \frac{\sum_{i=2}^{2^n}\lambda_i^k \left[\left(\overrightarrow{v_i},\overrightarrow{v_1}\right)_{S_{0}} \left(\overrightarrow{v_1},\overrightarrow{v_i}\right)_{S_{1}}\right]}{\lambda_1^k},
\end{equation}
\begin{equation}
    G_{\tau_0^m,\tau_{2}^{m+k}} = \frac{\sum_{i=2}^{2^n}\lambda_i^k \left[\left(\overrightarrow{v_i},\overrightarrow{v_1}\right)_{S_{0}} \left(\overrightarrow{v_1},\overrightarrow{v_i}\right)_{S_{2}}\right]}{\lambda_1^k}.
\end{equation}

	The spin matrices are
\begin{eqnarray*}
    S_0 = \operatorname{diag}[1,1,1,1,-1,-1,-1,-1];\\
    S_1 = \operatorname{diag}[1,1,-1,-1,1,1,-1,-1]; \\
    S_2 = \operatorname{diag}[1,-1,1,-1,1,-1,1,-1];
\end{eqnarray*}

\subsection{Hamiltonian with additional symmetry}
	Let the energy of our model not depend on the symmetries obtained as a result of reflections $\tau_i^m \leftrightarrow \tau_i^{m+1}$ where $i$ can take values from $\{0,1\}$, $\{0,2\}$, $\{1,2\}$. Then the transfer matrix has the form
\begin{equation}
    \Theta=
    \begin{pmatrix}
            a      &     b      &      b     &      c     &     b      &      c     &      c     &      d     \\ 
            b      &     e      &      f     &      g     &     f      &      g     &      i     &      j     \\
            b      &     f      &      e     &      g     &     f      &      i     &      g     &      j     \\
            c      &     g      &      g     &      k     &     i      &      l     &      l     &      m     \\
            b      &     f      &      f     &      i     &     e      &      g     &      g     &      j     \\
            c      &     g      &      i     &      l     &     g      &      k     &      l     &      m     \\
            c      &     i      &      g     &      l     &     g      &      l     &      k     &      m     \\
            d      &     j      &      j     &      m     &     j      &      m     &      m     &      n     
    \end{pmatrix}
\end{equation}
	where $g=h$ (taking the symmetry into account). In this case the partition function and correlation function have an analogous form, with allowance for $g=h$.
\subsubsection{Wilson loop}
	The matrix obtained with allowance for the interaction of the interlayer edges entering the Wilson loop contour of widths 1 and 2 has the form
\begin{equation}
    \Omega = 
\begin{pmatrix}
  \hat a&\hat b&\hat b&\hat c&\hat d&\hat f&\hat f&\hat g\\
  \hat b&\hat i&\hat c&\hat j&\hat f&\hat h&\hat g&\hat k\\
  \hat b&\hat c&\hat i&\hat j&\hat f&\hat g&\hat h&\hat k\\
  \hat c&\hat j&\hat j&\hat l&\hat g&\hat k&\hat k&\hat m\\
  \hat d&\hat f&\hat f&\hat g&\hat i&\hat j&\hat j&\hat n\\
  \hat f&\hat h&\hat g&\hat k&\hat j&\hat l&\hat n&\hat o\\
  \hat f&\hat g&\hat h&\hat k&\hat j&\hat n&\hat l&\hat o\\
  \hat g&\hat k&\hat k&\hat m&\hat n&\hat o&\hat o&\hat p
\end{pmatrix}
\end{equation}
	The eigenvalues of the matrix $\Omega$ are roots of a sixth-degree polynomial and a second-degree polynomial:
\begingroup
\footnotesize
    \begin{equation}
    \lambda_{7,8}=\frac{\hat c-\hat i+\hat l-\hat n \pm \sqrt{(\hat c-\hat i+\hat l-\hat n)^2-4(\hat i-\hat c)(\hat n-\hat l)-4(\hat h-\hat g)^2}}{2}
\end{equation}
\endgroup
	Its diagonalizing matrix has the form (with norms $M_i$ of the corresponding eigenvectors; in the case of coinciding eigenvalues we perform the standard Gram--Schmidt orthogonalization):
\begin{equation}
    R = 
\begin{pmatrix}
  \frac{r_{11}}{M_1}&\frac{r_{12}}{M_2}&\frac{r_{13}}{M_3}&\frac{r_{14}}{M_4}&\frac{r_{15}}{M_5}&\frac{r_{16}}{M_6}&0               &0               \\
  \frac{r_{21}}{M_1}&\frac{r_{22}}{M_2}&\frac{r_{23}}{M_3}&\frac{r_{24}}{M_4}&\frac{r_{25}}{M_5}&\frac{r_{26}}{M_6}&\frac{r_{27}}{M_7} &\frac{r_{28}}{M_8} \\
  \frac{r_{21}}{M_1}&\frac{r_{22}}{M_2}&\frac{r_{23}}{M_3}&\frac{r_{24}}{M_4}&\frac{r_{25}}{M_5}&\frac{r_{26}}{M_6}&-\frac{r_{27}}{M_7}&-\frac{r_{28}}{M_8}\\
  \frac{r_{41}}{M_1}&\frac{r_{42}}{M_2}&\frac{r_{43}}{M_3}&\frac{r_{44}}{M_4}&\frac{r_{45}}{M_5}&\frac{r_{46}}{M_6}&0               &0               \\
  \frac{r_{51}}{M_1}&\frac{r_{52}}{M_2}&\frac{r_{53}}{M_3}&\frac{r_{54}}{M_4}&\frac{r_{55}}{M_5}&\frac{r_{56}}{M_6}&0               &0               \\
  \frac{r_{61}}{M_1}&\frac{r_{62}}{M_2}&\frac{r_{63}}{M_3}&\frac{r_{64}}{M_4}&\frac{r_{65}}{M_5}&\frac{r_{66}}{M_6}&\frac{r_{67}}{M_7} &\frac{r_{68}}{M_8} \\
  \frac{r_{61}}{M_1}&\frac{r_{62}}{M_2}&\frac{r_{63}}{M_3}&\frac{r_{64}}{M_4}&\frac{r_{65}}{M_5}&\frac{r_{66}}{M_6}&-\frac{r_{67}}{M_7}&-\frac{r_{68}}{M_8}\\
  \frac{r_{81}}{M_1}&\frac{r_{72}}{M_2}&\frac{r_{83}}{M_3}&\frac{r_{84}}{M_4}&\frac{r_{85}}{M_5}&\frac{r_{86}}{M_6}&0               &0
\end{pmatrix},
\end{equation}

	The diagonalizing matrix of $\Theta$ has the form taken from Ref.~\cite{Volkov_Khrapov}. 
\begin{equation}
    Q = 
    \begin{pmatrix}
        \frac{\alpha_1}{N_1} &\frac{\alpha_2}{N_2} &\frac{\alpha_3}{N_3} &\frac{\alpha_4}{N_4} &0                                &0                                &0                                &0                                \\
        \frac{\beta_1}{N_1}  &\frac{\beta_2}{N_2}  &\frac{\beta_3}{N_3}  &\frac{\beta_4}{N_4}  &\frac{\omega \nu_5}{N_5}         &\frac{\omega \nu_6}{N_6}         &\frac{\overline\omega\nu_7}{N_7} &\frac{\overline\omega\nu_8}{N_8} \\
        \frac{\beta_1}{N_1}  &\frac{\beta_2}{N_2}  &\frac{\beta_3}{N_3}  &\frac{\beta_4}{N_4}  &\frac{\overline\omega\nu_5}{N_5} &\frac{\overline\omega\nu_6}{N_6} &\frac{\omega \nu_7}{N_7}         &\frac{\omega \nu_8}{N_8}         \\
        \frac{\gamma_1}{N_1} &\frac{\gamma_2}{N_2} &\frac{\gamma_3}{N_3} &\frac{\gamma_4}{N_4} &\frac{\overline\omega}{N_5}      &\frac{\overline\omega}{N_6}      &\frac{\omega}{N_7}               &\frac{\omega}{N_8}               \\
        \frac{\beta_1}{N_1}  &\frac{\beta_2}{N_2}  &\frac{\beta_3}{N_3}  &\frac{\beta_4}{N_4}  &\frac{\nu_5}{N_5}                &\frac{\nu_6}{N_6}                &\frac{\nu_7}{N_7}                &\frac{\nu_8}{N_8}                \\
        \frac{\gamma_1}{N_1} &\frac{\gamma_2}{N_2} &\frac{\gamma_3}{N_3} &\frac{\gamma_4}{N_4} &\frac{\omega}{N_5}               &\frac{\omega}{N_6}               &\frac{\overline\omega}{N_7}      &\frac{\overline\omega}{N_8}      \\
        \frac{\gamma_1}{N_1} &\frac{\gamma_2}{N_2} &\frac{\gamma_3}{N_3} &\frac{\gamma_4}{N_4} &\frac{1}{N_5}                    &\frac{1}{N_5}                    &\frac{1}{N_7}                    &\frac{1}{N_8}                    \\
        \frac{\delta_1}{N_1} &\frac{\delta_2}{N_2} &\frac{\delta_3}{N_3} &\frac{\delta_4}{N_4} &0                                &0                                &0                                &0
    \end{pmatrix},
\end{equation}
where $\omega= e^{\frac{2\pi}{3}i}$ and
\begin{eqnarray*}
    \nu_s = \frac{g+\omega g + i \overline \omega}{f - e + \lambda_s},& \qquad s=5,6;\\
    \nu_s = \frac{g+\overline\omega g + i \omega}{f - e + \lambda_s},& \qquad s=7,8;
\end{eqnarray*}
The spin matrices are
\begin{eqnarray*}
    S_\alpha = \operatorname{diag}[1,1,1,1,-1,-1,-1,-1];\\
    S_\beta = \operatorname{diag}[1,1,-1,-1,1,1,-1,-1]; \\
    S_\gamma = \operatorname{diag}[1,-1,1,-1,1,-1,1,-1];
\end{eqnarray*}
\begin{theorem}

		Explicit expression for the Wilson loop of the cyclically closed and torus-closed model in the thermodynamic limit ($L\to\infty$), covering a strip of width 1:
\begingroup
\small
        \begin{equation}\label{eq:3-chain Wilson Loop 1 enclosed}
    \begin{aligned}
    W_{i,j}&=\frac{Tr\{S_{\alpha} \Omega^{|j-i|}S_{\alpha} \Theta^{L-|j-i|}\}}{Z_L}=\\
    = &\frac{\mu_1^k}{\lambda_1^k}\left(\frac{\alpha_1 r_{11}+2\beta_1 r_{21}+\gamma_1 r_{41} - \beta_1 r_{51} - 2\gamma_1 r_{61} - \delta_1 r_{81}}{N_1 M_1}\right)^2 + \\
    + &\frac{\mu_2^k}{\lambda_1^k}\left(\frac{\alpha_1 r_{12}+2\beta_1 r_{22}+\gamma_1 r_{42} - \beta_1 r_{52} - 2\gamma_1 r_{62} - \delta_1 r_{82}}{N_1 M_2}\right)^2 + \\
    + &\frac{\mu_3^k}{\lambda_1^k}\left(\frac{\alpha_1 r_{13}+2\beta_1 r_{23}+\gamma_1 r_{43} - \beta_1 r_{53} - 2\gamma_1 r_{63} - \delta_1 r_{83}}{N_1 M_3}\right)^2 + \\
    + &\frac{\mu_4^k}{\lambda_1^k}\left(\frac{\alpha_1 r_{14}+2\beta_1 r_{24}+\gamma_1 r_{44} - \beta_1 r_{54} - 2\gamma_1 r_{64} - \delta_1 r_{84}}{N_1 M_4}\right)^2 + \\
    + &\frac{\mu_5^k}{\lambda_1^k}\left(\frac{\alpha_1 r_{15}+2\beta_1 r_{25}+\gamma_1 r_{45} - \beta_1 r_{55} - 2\gamma_1 r_{65} - \delta_1 r_{85}}{N_1 M_5}\right)^2 + \\
    + &\frac{\mu_6^k}{\lambda_1^k}\left(\frac{\alpha_1 r_{16}+2\beta_1 r_{26}+\gamma_1 r_{46} - \beta_1 r_{56} - 2\gamma_1 r_{66} - \delta_1 r_{86}}{N_1 M_6}\right)^2 
    \end{aligned}
\end{equation}
\endgroup
\end{theorem}
\begin{theorem}
		Explicit expression for the Wilson loop of the cyclically closed and torus-closed model in the thermodynamic limit ($L\to\infty$), covering a strip of width 2:
\begingroup
\small
        \begin{equation}\label{eq:3-chain Wilson Loop 2 enclosed}
    \begin{aligned}
    W_{i,j}&=\frac{Tr\{S_{\beta}S_{\gamma} \Omega^{|j-i|}S_{\beta}S_{\gamma} \Theta^{L-|j-i|}\}}{Z_L}=\\
    = &\frac{\mu_1^k}{\lambda_1^k}\left(\frac{\alpha_1 r_{11}-2\beta_1 r_{21}+\gamma_1 r_{41} + \beta_1 r_{51} - 2\gamma_1 r_{61} + \delta_1 r_{81}}{N_1 M_1}\right)^2 + \\
    + &\frac{\mu_2^k}{\lambda_1^k}\left(\frac{\alpha_1 r_{12}-2\beta_1 r_{22}+\gamma_1 r_{42} + \beta_1 r_{52} - 2\gamma_1 r_{62} + \delta_1 r_{82}}{N_1 M_2}\right)^2 + \\
    + &\frac{\mu_3^k}{\lambda_1^k}\left(\frac{\alpha_1 r_{13}-2\beta_1 r_{23}+\gamma_1 r_{43} + \beta_1 r_{53} - 2\gamma_1 r_{63} + \delta_1 r_{83}}{N_1 M_3}\right)^2 + \\
    + &\frac{\mu_4^k}{\lambda_1^k}\left(\frac{\alpha_1 r_{14}-2\beta_1 r_{24}+\gamma_1 r_{44} + \beta_1 r_{54} - 2\gamma_1 r_{64} + \delta_1 r_{84}}{N_1 M_4}\right)^2 + \\
    + &\frac{\mu_5^k}{\lambda_1^k}\left(\frac{\alpha_1 r_{15}-2\beta_1 r_{25}+\gamma_1 r_{45} + \beta_1 r_{55} - 2\gamma_1 r_{65} + \delta_1 r_{85}}{N_1 M_5}\right)^2 + \\
    + &\frac{\mu_6^k}{\lambda_1^k}\left(\frac{\alpha_1 r_{16}-2\beta_1 r_{26}+\gamma_1 r_{46} + \beta_1 r_{56} - 2\gamma_1 r_{66} + \delta_1 r_{86}}{N_1 M_6}\right)^2 
    \end{aligned}
\end{equation}
\endgroup
\end{theorem}
\subsection{String tension}
	Consider the Hamiltonian of the form
\begin{equation}\label{eq: triangle hamiltonian}
    \begin{aligned}
    \mathcal{H}_{n,m}^1 =& -\beta_l^*\left(\frac{1}{2}\sum_{\tau \in \{\tau_i^{m}, \tau_i^{m+1}\}}\tau + \sum_{\gamma \in \{\gamma^m\}}\gamma\right) - \\ &-\beta_t^*\sum_{t}\tau_t - \frac{1}{2}\alpha^*\left(\tau_0^m\tau_1^m\tau_2^m + \tau_0^{m+1} \tau_1^{m+1} \tau_2^{m+1}\right)
    \end{aligned}
\end{equation}
	where $\tau_t$ are products of spins in the triangles shown in Fig.~\ref{fig: three-chain transform} (the coefficient $1/2$ is related to taking into account the interactions on the neighboring layers with $m$). We divide the given Hamiltonian by $\alpha^*>0$ and pass to the parameters $\beta_l$, $\beta_t$.

	We now use expression~\eqref{eq: Wilson n infty}. If we take the limit $k\to\infty$ in the area-law region, then in the thermodynamic limit for sufficiently large contours the Wilson loop is equivalent to a single term:
\begin{equation}
    WL_k^d \sim \left(\frac{\mu_1}{\lambda_1}\right)^k  \left|\left(\overrightarrow{r_1},{\overrightarrow{v_1}}\right)_{S_{\left[\alpha_1\alpha_d\right]}}\right|^2 \sim e^{-\sigma S}
\end{equation}
	From this one can express the string tension for $k\to\infty$:
\begin{equation}\label{eq: string}
    \sigma = -\frac{1}{d}\ln\frac{\mu_1}{\lambda_1}
\end{equation}
	It is known that the string tension tends to a nonzero constant in the given limit. Also, by abandoning the symmetry in the Hamiltonian~\eqref{eq: triangle hamiltonian}, for example by keeping only the triangles with hypotenuses $\gamma_{0,1}^m$, $\gamma_{1,2}^m$, $\gamma_{2,3}^m$, we lose the symmetry of the Hamiltonian but obtain a similar result.
\begin{figure}[htbp]
  \centering
  \begin{subfigure}[b]{0.45\textwidth}
    \centering
    \includegraphics[width=\textwidth]{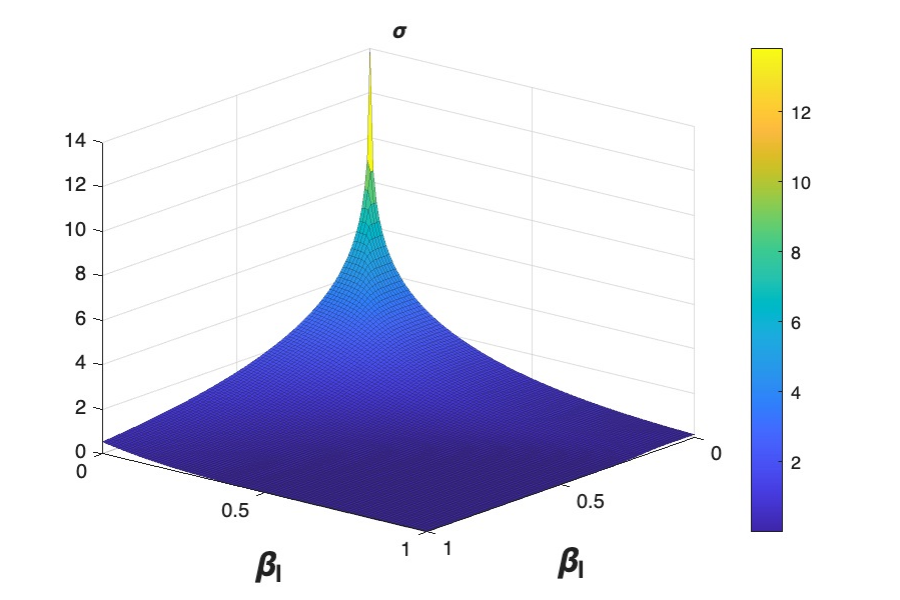}
    \caption{}
    \label{fig:surface with double triangles}
  \end{subfigure}
  \hfill 
  \begin{subfigure}[b]{0.45\textwidth}
    \centering
    \includegraphics[width=\textwidth]{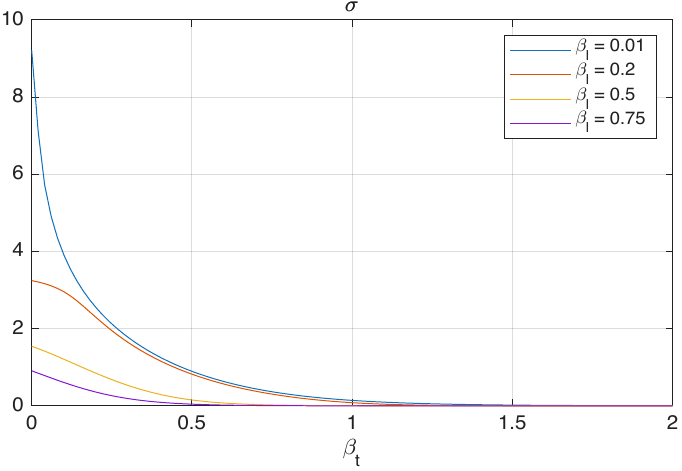}
    \caption{}
    \label{fig:plots with double triangles}
  \end{subfigure}
  \caption{String tension \eqref{fig:surface with double triangles} hamiltonian  \eqref{eq: triangle hamiltonian}   with $\beta_l\in\left[0.001,1\right]$ and $\beta_t\in\left[0,1\right]$ and plots \eqref{fig:plots with double triangles} with $\beta_t\in\left[0,1\right], \beta_l = 0.01, 0.2, 0.5, 0.75$.}
  \label{fig:model with double triangles}
\end{figure}
 Also, by abandoning the symmetry in the Hamiltonian~\eqref{eq: triangle hamiltonian}, for example by keeping only the triangles with hypotenuses $\gamma_{0,1}^m$, $\gamma_{1,2}^m$, $\gamma_{2,3}^m$, we lose the symmetry of the Hamiltonian but obtain a similar result.\begin{figure}[htbp]
  \centering
  \begin{subfigure}[b]{0.45\textwidth}
    \centering
    \includegraphics[width=\textwidth]{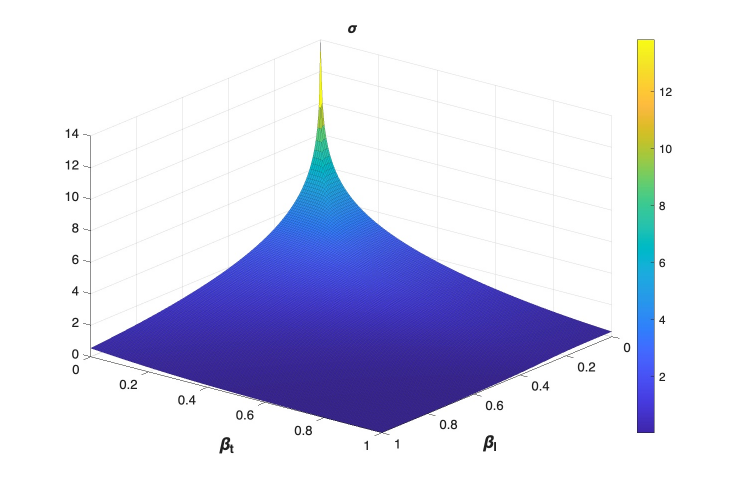}
    \caption{}
    \label{fig:surface with solo triangles}
  \end{subfigure}
  \hfill 
  \begin{subfigure}[b]{0.45\textwidth}
    \centering
    \includegraphics[width=\textwidth]{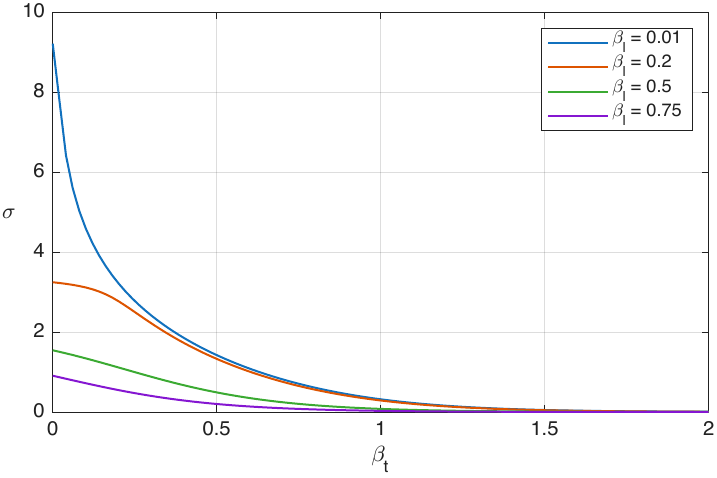}
    \caption{}
    \label{fig:plots with solo triangles}
  \end{subfigure}
  \caption{String tension \eqref{fig:surface with double triangles} with hamiltonian  \eqref{eq: triangle hamiltonian}, without symmetry, with $\beta_l\in\left[0.001,1\right]$ and $\beta_t\in\left[0,1\right]$ and plots \eqref{fig:plots with double triangles} with $\beta_t\in\left[0,1\right], \beta_l = 0.01, 0.2, 0.5, 0.75$.}
  \label{fig:model with solo triangles}
\end{figure}

\section{Four-Chain Model}\label{sec:n4}
\subsection{Model with plaquette and nearest-neighbour interactions}
\begin{equation}
\Theta = 
\begin{pmatrix}
a & b & b & c & b & d & c & h & b & c & d & h & c & h & h & f \\
b & v & i & j & k & j & h & c & i & j & h & d & h & c & l & m\\
b & i & v & j & i & h & j & d & k & h & j & c & h & l & c & m\\
c & j & j & n & h & o & p & q & h & p & o & q & f & m & m & r\\
b & k & i & h & v & j & j & c & i & h & h & l & j & c & d & m\\
d & j & h & o & j & n & o & q & h & o & f & m & o & q & m & s\\
c & h & j & p & j & o & n & q & h & f & o & m & p & m & q & r\\
h & c & d & q & c & q & q & n & l & m & m & o & m & p & o & t\\
b & i & k & h & i & h & h & l & v & j & j & c & j & d & c & m\\
c & j & h & p & h & o & f & m & j & n & o & q & p & q & m & r\\
d & h & j & o & h & f & o & m & j & o & n & q & o & m & q & s\\
h & d & c & q & l & m & m & o & c & q & q & n & m & o & p & t\\
c & h & h & f & j & o & p & m & j & p & o & m & n & q & q & r\\
h & c & l & m & c & q & m & p & d & q & m & o & q & n & o & t\\
h & l & c & m & d & m & q & o & c & m & q & p & q & o & n & t\\
f & m & m & r & m & s & r & t & m & r & s & t & r & t & t & u
\end{pmatrix}
\end{equation}
The transfer matrix commutes with the matrix displayed in Eq. \eqref{eq: permutation 4-chain}; consequently its 16 eigenvalues and eigenvectors fall into well-defined irreducible representations of the underlying symmetry.
\begin{widetext}
\begingroup
\footnotesize
\begin{equation}\label{eq: permutation 4-chain}
    \begin{pmatrix}
1 & 0 & 0 & 0 & 0 & 0 & 0 & 0 & 0 & 0 & 0 & 0 & 0 & 0 & 0 \\
0 & 0 & 1 & 0 & 0 & 0 & 0 & 0 & 0 & 0 & 0 & 0 & 0 & 0 & 0 \\
0 & 0 & 0 & 0 & 1 & 0 & 0 & 0 & 0 & 0 & 0 & 0 & 0 & 0 & 0 \\
0 & 0 & 0 & 0 & 0 & 0 & 1 & 0 & 0 & 0 & 0 & 0 & 0 & 0 & 0 \\
0 & 0 & 0 & 0 & 0 & 0 & 0 & 0 & 1 & 0 & 0 & 0 & 0 & 0 & 0 \\
0 & 0 & 0 & 0 & 0 & 0 & 0 & 0 & 0 & 0 & 1 & 0 & 0 & 0 & 0 \\
0 & 0 & 0 & 0 & 0 & 0 & 0 & 0 & 0 & 0 & 0 & 0 & 1 & 0 & 0 \\
0 & 0 & 0 & 0 & 0 & 0 & 0 & 0 & 0 & 0 & 0 & 0 & 0 & 0 & 1 \\
0 & 1 & 0 & 0 & 0 & 0 & 0 & 0 & 0 & 0 & 0 & 0 & 0 & 0 & 0 \\
0 & 0 & 0 & 1 & 0 & 0 & 0 & 0 & 0 & 0 & 0 & 0 & 0 & 0 & 0 \\
0 & 0 & 0 & 0 & 0 & 1 & 0 & 0 & 0 & 0 & 0 & 0 & 0 & 0 & 0 \\
0 & 0 & 0 & 0 & 0 & 0 & 0 & 1 & 0 & 0 & 0 & 0 & 0 & 0 & 0 \\
0 & 0 & 0 & 0 & 0 & 0 & 0 & 0 & 0 & 1 & 0 & 0 & 0 & 0 & 0 \\
0 & 0 & 0 & 0 & 0 & 0 & 0 & 0 & 0 & 0 & 0 & 1 & 0 & 0 & 0 \\
0 & 0 & 0 & 0 & 0 & 0 & 0 & 0 & 0 & 0 & 0 & 0 & 0 & 1 & 0 \\
0 & 0 & 0 & 0 & 0 & 0 & 0 & 0 & 0 & 0 & 0 & 0 & 0 & 0 & 1
\end{pmatrix}
\end{equation}
\endgroup
 This symmetry matrix has eigenvalues $+i$ and $-i$, each with algebraic multiplicity three.
 The diagonalizing matrix $Q$  consists of orthonormalized and normalized eigenvectors (real elements):

\begin{equation}
    Q =
    \begin{pmatrix}
\frac{\alpha_1}{N_1} & 0 & 0 & 0 & 0 & 0 & \frac{\alpha_7}{N_7} & \frac{\alpha_8}{N_8} & 0 & \frac{\alpha_{10}}{N_{10}} & 0 & 0 & 0 & 0 & \frac{\alpha_{15}}{N_{15}} & \frac{\alpha_{16}}{N_{16}} \\
\frac{\beta_1}{N_1} & \frac{\alpha_2}{N_2} & \frac{\alpha_3}{N_3} & 0 & \frac{\alpha_5}{N_5} & 0 & \frac{\beta_7}{N_7} & \frac{\beta_8}{N_8} & \frac{\alpha_9}{N_9} & \frac{\beta_{10}}{N_{10}} & 0 & \frac{\alpha_{12}}{N_{12}} & \frac{\alpha_{13}}{N_{13}} & 0 & \frac{\beta_{15}}{N_{15}} & \frac{\beta_{16}}{N_{16}} \\
\frac{\beta_1}{N_1} & -\frac{\alpha_2}{N_2} & 0 & \frac{\alpha_3}{N_3} & 0 & \frac{\alpha_5}{N_5} & \frac{\beta_7}{N_7} & \frac{\beta_8}{N_8} & -\frac{\alpha_9}{N_9} & \frac{\beta_{10}}{N_{10}} & 0 & -\frac{\alpha_{12}}{N_{12}} & 0 & \frac{\alpha_{13}}{N_{13}} & \frac{\beta_{15}}{N_{15}} & \frac{\beta_{16}}{N_{16}} \\
\frac{\gamma_1}{N_1} & 0 & \frac{\beta_3}{N_3} & \frac{\beta_3}{N_3} & \frac{\beta_5}{N_5} & \frac{\beta_5}{N_5} & \frac{\gamma_7}{N_7} & \frac{\gamma_8}{N_8} & 0 & \frac{\gamma_{10}}{N_{10}} & \frac{1}{N_{11}} & 0 & \frac{\beta_{13}}{N_{13}} & \frac{\beta_{13}}{N_{13}} & \frac{\gamma_{15}}{N_{15}} & \frac{\gamma_{16}}{N_{16}} \\
\frac{\beta_1}{N_1} & \frac{\alpha_2}{N_2} & -\frac{\alpha_3}{N_3} & 0 & -\frac{\alpha_5}{N_5} & 0 & \frac{\beta_7}{N_7} & \frac{\beta_8}{N_8} & \frac{\alpha_9}{N_9} & \frac{\beta_{10}}{N_{10}} & 0 & \frac{\alpha_{12}}{N_{12}} & -\frac{\alpha_{13}}{N_{13}} & 0 & \frac{\beta_{15}}{N_{15}} & \frac{\beta_{16}}{N_{16}} \\
\frac{\delta_1}{N_1} & \frac{\beta_2}{N_2} & 0 & 0 & 0 & 0 & \frac{\delta_7}{N_7} & \frac{\delta_8}{N_8} & \frac{\beta_9}{N_9} & \frac{\delta_{10}}{N_{10}} & 0 & \frac{\beta_{12}}{N_{12}} & 0 & 0 & \frac{\delta_{15}}{N_{15}} & \frac{\delta_{16}}{N_{16}} \\
\frac{\gamma_1}{N_1} & 0 & -\frac{\beta_3}{N_3} & \frac{\beta_3}{N_3} & -\frac{\beta_5}{N_5} & \frac{\beta_5}{N_5} & \frac{\gamma_7}{N_7} & \frac{\gamma_8}{N_8} & 0 & \frac{\gamma_{10}}{N_{10}} & -\frac{1}{N_{11}} & 0 & -\frac{\beta_{13}}{N_{13}} & \frac{\beta_{13}}{N_{13}} & \frac{\gamma_{15}}{N_{15}} & \frac{\gamma_{16}}{N_{16}} \\
\frac{\varepsilon_1}{N_1} & -\frac{1}{N_2} & 0 & -\frac{1}{N_3} & 0 & -\frac{1}{N_5} & \frac{\varepsilon_7}{N_7} & \frac{\varepsilon_8}{N_8} & -\frac{1}{N_9} & \frac{\varepsilon_{10}}{N_{10}} & 0 & -\frac{1}{N_{12}} & 0 & -\frac{1}{N_{13}} & \frac{\varepsilon_{15}}{N_{15}} & \frac{\varepsilon_{16}}{N_{16}} \\
\frac{\beta_1}{N_1} & -\frac{\alpha_2}{N_2} & 0 & -\frac{\alpha_3}{N_3} & 0 & -\frac{\alpha_5}{N_5} & \frac{\beta_7}{N_7} & \frac{\beta_8}{N_8} & -\frac{\alpha_9}{N_9} & \frac{\beta_{10}}{N_{10}} & 0 & -\frac{\alpha_{12}}{N_{12}} & 0 & -\frac{\alpha_{13}}{N_{13}} & \frac{\beta_{15}}{N_{15}} & \frac{\beta_{16}}{N_{16}} \\
\frac{\gamma_1}{N_1} & 0 & \frac{\beta_3}{N_3} & -\frac{\beta_3}{N_3} & \frac{\beta_5}{N_5} & -\frac{\beta_5}{N_5} & \frac{\gamma_7}{N_7} & \frac{\gamma_8}{N_8} & 0 & \frac{\gamma_{10}}{N_{10}} & -\frac{1}{N_{11}} & 0 & \frac{\beta_{13}}{N_{13}} & -\frac{\beta_{13}}{N_{13}} & \frac{\gamma_{15}}{N_{15}} & \frac{\gamma_{16}}{N_{16}} \\
\frac{\delta_1}{N_1} & -\frac{\beta_2}{N_2} & 0 & 0 & 0 & 0 & \frac{\delta_7}{N_7} & \frac{\delta_8}{N_8} & -\frac{\beta_9}{N_9} & \frac{\delta_{10}}{N_{10}} & 0 & -\frac{\beta_{12}}{N_{12}} & 0 & 0 & \frac{\delta_{15}}{N_{15}} & \frac{\delta_{16}}{N_{16}} \\
\frac{\varepsilon_1}{N_1} & \frac{1}{N_2} & -\frac{1}{N_3} & 0 & -\frac{1}{N_5} & 0 & \frac{\varepsilon_7}{N_7} & \frac{\varepsilon_8}{N_8} & \frac{1}{N_9} & \frac{\varepsilon_{10}}{N_{10}} & 0 & \frac{1}{N_{12}} & -\frac{1}{N_{13}} & 0 & \frac{\varepsilon_{15}}{N_{15}} & \frac{\varepsilon_{16}}{N_{16}} \\
\frac{\gamma_1}{N_1} & 0 & -\frac{\beta_3}{N_3} & -\frac{\beta_3}{N_3} & -\frac{\beta_5}{N_5} & -\frac{\beta_5}{N_5} & \frac{\gamma_7}{N_7} & \frac{\gamma_8}{N_8} & 0 & \frac{\gamma_{10}}{N_{10}} & \frac{1}{N_{11}} & 0 & -\frac{\beta_{13}}{N_{13}} & -\frac{\beta_{13}}{N_{13}} & \frac{\gamma_{15}}{N_{15}} & \frac{\gamma_{16}}{N_{16}} \\
\frac{\varepsilon_1}{N_1} & -\frac{1}{N_2} & 0 & \frac{1}{N_3} & 0 & \frac{1}{N_5} & \frac{\varepsilon_7}{N_7} & \frac{\varepsilon_8}{N_8} & -\frac{1}{N_9} & \frac{\varepsilon_{10}}{N_{10}} & 0 & -\frac{1}{N_{12}} & 0 & \frac{1}{N_{13}} & \frac{\varepsilon_{15}}{N_{15}} & \frac{\varepsilon_{16}}{N_{16}} \\
\frac{\varepsilon_1}{N_1} & \frac{1}{N_2} & \frac{1}{N_3} & 0 & \frac{1}{N_5} & 0 & \frac{\varepsilon_7}{N_7} & \frac{\varepsilon_8}{N_8} & \frac{1}{N_9} & \frac{\varepsilon_{10}}{N_{10}} & 0 & \frac{1}{N_{12}} & \frac{1}{N_{13}} & 0 & \frac{\varepsilon_{15}}{N_{15}} & \frac{\varepsilon_{16}}{N_{16}} \\
\frac{1}{N_1} & 0 & 0 & 0 & 0 & 0 & \frac{1}{N_7} & \frac{1}{N_8} & 0 & \frac{1}{N_{10}} & 0 & 0 & 0 & 0 & \frac{1}{N_{15}} & \frac{1}{N_{16}}
    \end{pmatrix}
\end{equation}
\end{widetext}
If the matrix has more degenerate eigenvalues, we can apply the Gram-Schmidt orthogonalization and compute the correlation functions and Wilson loops using the same method or via the general formulas of the theorems described above \ref{sec: theorems}.
\begin{theorem}
		For the four-chain cyclically closed and torus-closed model in the thermodynamic limit ($L\to\infty$):
\begin{gather}
    \begin{align}
        &\left<\sigma_1^i\sigma_1^j\right>_2 - \left<\sigma_1^i\right>_2\left<\sigma_1^j\right>_2 = \frac{\sum_{i=2}^{16}\lambda_i^k \left[\left(\overrightarrow{v_i},\overrightarrow{v_1}\right)_{S_{1}} \left(\overrightarrow{v_1},\overrightarrow{v_i}\right)_{S_{1}}\right]}{\lambda_1^k},
    \end{align}
\end{gather}

\begin{gather}
    \begin{align}
        &\left<\sigma_1^i\sigma_2^j\right>_2 - \left<\sigma_1^i\right>_2\left<\sigma_2^j\right>_2 = \frac{\sum_{i=2}^{16}\lambda_i^k \left[\left(\overrightarrow{v_i},\overrightarrow{v_1}\right)_{S_{1}} \left(\overrightarrow{v_1},\overrightarrow{v_i}\right)_{S_{2}}\right]}{\lambda_1^k},
    \end{align}
\end{gather}

\begin{gather}
    \begin{align}
        &\left<\sigma_1^i\sigma_3^j\right>_2 - \left<\sigma_1^i\right>_2\left<\sigma_3^j\right>_2 = \frac{\sum_{i=2}^{16}\lambda_i^k \left[\left(\overrightarrow{v_i},\overrightarrow{v_1}\right)_{S_{1}} \left(\overrightarrow{v_1},\overrightarrow{v_i}\right)_{S_{3}}\right]}{\lambda_1^k},
    \end{align}
\end{gather}

\begin{gather}
    \begin{align}
        \left<\sigma_1^i\sigma_4^j\right>_2 - \left<\sigma_1^i\right>_2&\left<\sigma_4^j\right>_2 = \left<\sigma_1^i\sigma_2^j\right>_2 - \left<\sigma_1^i\right>_2\left<\sigma_2^j\right>_2,
    \end{align}
\end{gather}
where
\begin{equation}
    S_1 = \operatorname{diag}[1,1,1,1,1,1,1,1,-1,-1,-1,-1,-1,-1,-1,-1]
\end{equation}
\begin{equation}
    S_2 = \operatorname{diag}[1,1,1,1,-1,-1,-1,-1,1,1,1,1,-1,-1,-1,-1]
\end{equation}
\begin{equation}
    S_3 = \operatorname{diag}[1,1,-1,-1,1,1,-1,-1,1,1,-1,-1,1,1,-1,-1]
\end{equation}
\end{theorem}
		These expressions are valid for all sets of spins whose lower indices are subjected to the permutation $(1234)$ and its powers.
	
        The eigenvalues corresponding to the subspaces $i$ and $-i$ in matrix~\eqref{eq: permutation 4-chain} have a positive coefficient when we consider spins on one chain (rotation by a multiple of $2\pi$), a zero coefficient when considering spins on neighboring chains (rotation by $\pi/2$ --- zero real part, nonzero imaginary part, as $e^{i\pi/2}$), and a negative coefficient when considering spins on opposite chains (angle $\pi$, $\pm i^2=-1$).
        \begin{widetext}
\begingroup
\footnotesize
        \begin{equation}
    \Omega_1=
\begin{pmatrix}
\tilde{a} & \tilde{b} & \tilde{c} & \tilde{d} & \tilde{b} & \tilde{f} & \tilde{d} & \tilde{g} & \tilde{h} & \tilde{i} & \tilde{j} & \tilde{k} & \tilde{i} & \tilde{l} & \tilde{k} & \tilde{m} \\
\tilde{b} & \tilde{n} & \tilde{o} & \tilde{p} & \tilde{q} & \tilde{r} & \tilde{g} & \tilde{s} & \tilde{t} & \tilde{u} & \tilde{k} & \tilde{v} & \tilde{l} & \tilde{w} & \tilde{x} & \tilde{y} \\
\tilde{c} & \tilde{o} & \tilde{n} & \tilde{r} & \tilde{o} & \tilde{g} & \tilde{r} & \tilde{z} & \tilde{\alpha} & \tilde{k} & \tilde{u} & \tilde{w} & \tilde{k} & \tilde{x} & \tilde{w} & \tilde{\beta} \\
\tilde{d} & \tilde{p} & \tilde{r} & \tilde{\gamma} & \tilde{g} & \tilde{\delta} & \tilde{\varphi} & \tilde{\psi} & \tilde{k} & \tilde{\eta} & \tilde{\psi} & \tilde{\iota} & \tilde{m} & \tilde{y} & \tilde{\beta} & \tilde{\vartheta} \\
\tilde{b} & \tilde{q} & \tilde{o} & \tilde{g} & \tilde{n} & \tilde{r} & \tilde{p} & \tilde{s} & \tilde{t} & \tilde{l} & \tilde{k} & \tilde{x} & \tilde{u} & \tilde{w} & \tilde{v} & \tilde{y} \\
\tilde{f} & \tilde{r} & \tilde{g} & \tilde{\delta} & \tilde{r} & \tilde{\gamma} & \tilde{\delta} & \tilde{\kappa} & \tilde{l} & \tilde{\psi} & \tilde{m} & \tilde{y} & \tilde{\psi} & \tilde{\iota} & \tilde{y} & \tilde{\lambda} \\
\tilde{d} & \tilde{g} & \tilde{r} & \tilde{\varphi} & \tilde{p} & \tilde{\delta} & \tilde{\gamma} & \tilde{\psi} & \tilde{k} & \tilde{m} & \tilde{\psi} & \tilde{\beta} & \tilde{\eta} & \tilde{y} & \tilde{\iota} & \tilde{\vartheta} \\
\tilde{g} & \tilde{s} & \tilde{z} & \tilde{\psi} & \tilde{s} & \tilde{\kappa} & \tilde{\psi} & \tilde{\mu} & \tilde{x} & \tilde{y} & \tilde{\beta} & \tilde{\nu} & \tilde{y} & \tilde{\theta} & \tilde{\nu} & \tilde{\pi} \\
\tilde{h} & \tilde{t} & \tilde{\alpha} & \tilde{k} & \tilde{t} & \tilde{l} & \tilde{k} & \tilde{x} & \tilde{n} & \tilde{r} & \tilde{p} & \tilde{s} & \tilde{r} & \tilde{z} & \tilde{s} & \tilde{\rho} \\
\tilde{i} & \tilde{u} & \tilde{k} & \tilde{\eta} & \tilde{l} & \tilde{\psi} & \tilde{m} & \tilde{y} & \tilde{r} & \tilde{\gamma} & \tilde{\delta} & \tilde{\kappa} & \tilde{\varphi} & \tilde{\psi} & \tilde{\rho} & \tilde{\sigma} \\
\tilde{j} & \tilde{k} & \tilde{u} & \tilde{\psi} & \tilde{k} & \tilde{m} & \tilde{\psi} & \tilde{\beta} & \tilde{p} & \tilde{\delta} & \tilde{\gamma} & \tilde{\psi} & \tilde{\delta} & \tilde{\rho} & \tilde{\psi} & \tilde{\varsigma} \\
\tilde{k} & \tilde{v} & \tilde{w} & \tilde{\iota} & \tilde{x} & \tilde{y} & \tilde{\beta} & \tilde{\nu} & \tilde{s} & \tilde{\kappa} & \tilde{\psi} & \tilde{\mu} & \tilde{\rho} & \tilde{\tau} & \tilde{\upsilon} & \tilde{\chi} \\
\tilde{i} & \tilde{l} & \tilde{k} & \tilde{m} & \tilde{u} & \tilde{\psi} & \tilde{\eta} & \tilde{y} & \tilde{r} & \tilde{\varphi} & \tilde{\delta} & \tilde{\rho} & \tilde{\gamma} & \tilde{\psi} & \tilde{\kappa} & \tilde{\sigma} \\
\tilde{l} & \tilde{w} & \tilde{x} & \tilde{y} & \tilde{w} & \tilde{\iota} & \tilde{y} & \tilde{\theta} & \tilde{z} & \tilde{\psi} & \tilde{\rho} & \tilde{\tau} & \tilde{\psi} & \tilde{\mu} & \tilde{\tau} & \tilde{\omega} \\
\tilde{k} & \tilde{x} & \tilde{w} & \tilde{\beta} & \tilde{v} & \tilde{y} & \tilde{\iota} & \tilde{\nu} & \tilde{s} & \tilde{\rho} & \tilde{\psi} & \tilde{\upsilon} & \tilde{\kappa} & \tilde{\tau} & \tilde{\mu} & \tilde{\chi} \\
\tilde{m} & \tilde{y} & \tilde{\beta} & \tilde{\vartheta} & \tilde{y} & \tilde{\lambda} & \tilde{\vartheta} & \tilde{\pi} & \tilde{\rho} & \tilde{\sigma} & \tilde{\varsigma} & \tilde{\chi} & \tilde{\sigma} & \tilde{\omega} & \tilde{\chi} & \tilde{\xi}
\end{pmatrix}
\end{equation}

\begin{equation}
    \Omega_2=
\begin{pmatrix}
\tilde{a} & \tilde{b} & \tilde{b} & \tilde{c} & \tilde{b} & \tilde{d} & \tilde{f} & \tilde{g} & \tilde{b} & \tilde{f} & \tilde{d} & \tilde{g} & \tilde{c} & \tilde{g} & \tilde{g} & \tilde{h} \\
\tilde{b} & \tilde{i} & \tilde{j} & \tilde{k} & \tilde{l} & \tilde{k} & \tilde{g} & \tilde{m} & \tilde{n} & \tilde{k} & \tilde{g} & \tilde{o} & \tilde{g} & \tilde{p} & \tilde{q} & \tilde{r} \\
\tilde{b} & \tilde{j} & \tilde{i} & \tilde{k} & \tilde{n} & \tilde{g} & \tilde{k} & \tilde{o} & \tilde{l} & \tilde{g} & \tilde{k} & \tilde{m} & \tilde{g} & \tilde{q} & \tilde{p} & \tilde{r} \\
\tilde{c} & \tilde{k} & \tilde{k} & \tilde{s} & \tilde{g} & \tilde{t} & \tilde{u} & \tilde{v} & \tilde{g} & \tilde{u} & \tilde{t} & \tilde{v} & \tilde{h} & \tilde{r} & \tilde{r} & \tilde{w} \\
\tilde{b} & \tilde{l} & \tilde{n} & \tilde{g} & \tilde{i} & \tilde{k} & \tilde{k} & \tilde{p} & \tilde{j} & \tilde{g} & \tilde{g} & \tilde{q} & \tilde{k} & \tilde{m} & \tilde{o} & \tilde{r} \\
\tilde{d} & \tilde{k} & \tilde{g} & \tilde{t} & \tilde{k} & \tilde{s} & \tilde{x} & \tilde{v} & \tilde{g} & \tilde{x} & \tilde{h} & \tilde{r} & \tilde{t} & \tilde{v} & \tilde{r} & \tilde{y} \\
\tilde{f} & \tilde{g} & \tilde{k} & \tilde{u} & \tilde{k} & \tilde{x} & \tilde{s} & \tilde{v} & \tilde{g} & \tilde{h} & \tilde{x} & \tilde{r} & \tilde{u} & \tilde{r} & \tilde{v} & \tilde{z} \\
\tilde{g} & \tilde{m} & \tilde{o} & \tilde{v} & \tilde{p} & \tilde{v} & \tilde{v} & \tilde{\alpha} & \tilde{q} & \tilde{r} & \tilde{r} & \tilde{\beta} & \tilde{r} & \tilde{\gamma} & \tilde{\delta} & \tilde{\varepsilon} \\
\tilde{b} & \tilde{n} & \tilde{l} & \tilde{g} & \tilde{j} & \tilde{g} & \tilde{g} & \tilde{q} & \tilde{i} & \tilde{k} & \tilde{k} & \tilde{p} & \tilde{k} & \tilde{o} & \tilde{m} & \tilde{r} \\
\tilde{f} & \tilde{k} & \tilde{g} & \tilde{u} & \tilde{g} & \tilde{x} & \tilde{h} & \tilde{r} & \tilde{k} & \tilde{s} & \tilde{x} & \tilde{v} & \tilde{u} & \tilde{v} & \tilde{r} & \tilde{z} \\
\tilde{d} & \tilde{g} & \tilde{k} & \tilde{t} & \tilde{g} & \tilde{h} & \tilde{x} & \tilde{r} & \tilde{k} & \tilde{x} & \tilde{s} & \tilde{v} & \tilde{t} & \tilde{r} & \tilde{v} & \tilde{y} \\
\tilde{g} & \tilde{o} & \tilde{m} & \tilde{v} & \tilde{q} & \tilde{r} & \tilde{r} & \tilde{\beta} & \tilde{p} & \tilde{v} & \tilde{v} & \tilde{\alpha} & \tilde{r} & \tilde{\delta} & \tilde{\gamma} & \tilde{\varepsilon} \\
\tilde{c} & \tilde{g} & \tilde{g} & \tilde{h} & \tilde{k} & \tilde{t} & \tilde{u} & \tilde{r} & \tilde{k} & \tilde{u} & \tilde{t} & \tilde{r} & \tilde{s} & \tilde{v} & \tilde{v} & \tilde{w} \\
\tilde{g} & \tilde{p} & \tilde{q} & \tilde{r} & \tilde{m} & \tilde{v} & \tilde{r} & \tilde{\gamma} & \tilde{o} & \tilde{v} & \tilde{r} & \tilde{\delta} & \tilde{v} & \tilde{\alpha} & \tilde{\beta} & \tilde{\varepsilon} \\
\tilde{g} & \tilde{q} & \tilde{p} & \tilde{r} & \tilde{o} & \tilde{r} & \tilde{v} & \tilde{\delta} & \tilde{m} & \tilde{r} & \tilde{v} & \tilde{\gamma} & \tilde{v} & \tilde{\beta} & \tilde{\alpha} & \tilde{\varepsilon} \\
\tilde{h} & \tilde{r} & \tilde{r} & \tilde{w} & \tilde{r} & \tilde{y} & \tilde{z} & \tilde{\varepsilon} & \tilde{r} & \tilde{z} & \tilde{y} & \tilde{\varepsilon} & \tilde{w} & \tilde{\varepsilon} & \tilde{\varepsilon} & \tilde{\zeta}
\end{pmatrix}
\end{equation}
\begin{equation}
    \Omega_3=\Omega_1
\end{equation}
Diagonalizing matrix for $\Omega_1$:
\begin{equation}
\begin{pmatrix}
\tilde{\alpha}_1 & \tilde{\alpha}_2 & \tilde{\alpha}_3 & \tilde{\alpha}_4 & 0 & \tilde{\alpha}_6 & 0 & \tilde{\alpha}_8 & \tilde{\alpha}_9 & \tilde{\alpha}_{10} & \tilde{\alpha}_{11} & \tilde{\alpha}_{12} & 0 & \tilde{\alpha}_{14} & 0 & \tilde{\alpha}_{16} \\
\tilde{\beta}_1 & \tilde{\beta}_2 & \tilde{\beta}_3 & \tilde{\beta}_4 & \tilde{\alpha}_5 & \tilde{\beta}_6 & \tilde{\alpha}_7 & \tilde{\beta}_8 & \tilde{\beta}_9 & \tilde{\beta}_{10} & \tilde{\beta}_{11} & \tilde{\beta}_{12} & \tilde{\alpha}_{13} & \tilde{\beta}_{14} & \tilde{\alpha}_{15} & \tilde{\beta}_{16} \\
\tilde{\gamma}_1 & \tilde{\gamma}_2 & \tilde{\gamma}_3 & \tilde{\gamma}_4 & 0 & \tilde{\gamma}_6 & 0 & \tilde{\gamma}_8 & \tilde{\gamma}_9 & \tilde{\gamma}_{10} & \tilde{\gamma}_{11} & \tilde{\gamma}_{12} & 0 & \tilde{\gamma}_{14} & 0 & \tilde{\gamma}_{16} \\
\tilde{\delta}_1 & \tilde{\delta}_2 & \tilde{\delta}_3 & \tilde{\delta}_4 & -\tilde{\beta}_5 & \tilde{\delta}_6 & -\tilde{\beta}_7 & \tilde{\delta}_8 & \tilde{\delta}_9 & \tilde{\delta}_{10} & \tilde{\delta}_{11} & \tilde{\delta}_{12} & -\tilde{\beta}_{13} & \tilde{\delta}_{14} & -\tilde{\beta}_{15} & \tilde{\delta}_{16} \\
\tilde{\beta}_1 & \tilde{\beta}_2 & \tilde{\beta}_3 & \tilde{\beta}_4 & -\tilde{\alpha}_5 & \tilde{\beta}_6 & -\tilde{\alpha}_7 & \tilde{\beta}_8 & \tilde{\beta}_9 & \tilde{\beta}_{10} & \tilde{\beta}_{11} & \tilde{\beta}_{12} & -\tilde{\alpha}_{13} & \tilde{\beta}_{14} & -\tilde{\alpha}_{15} & \tilde{\beta}_{16} \\
\tilde{\varepsilon}_1 & \tilde{\varepsilon}_2 & \tilde{\varepsilon}_3 & \tilde{\varepsilon}_4 & 0 & \tilde{\varepsilon}_6 & 0 & \tilde{\varepsilon}_8 & \tilde{\varepsilon}_9 & \tilde{\varepsilon}_{10} & \tilde{\varepsilon}_{11} & \tilde{\varepsilon}_{12} & 0 & \tilde{\varepsilon}_{14} & 0 & \tilde{\varepsilon}_{16} \\
\tilde{\delta}_1 & \tilde{\delta}_2 & \tilde{\delta}_3 & \tilde{\delta}_4 & \tilde{\beta}_5 & \tilde{\delta}_6 & \tilde{\beta}_7 & \tilde{\delta}_8 & \tilde{\delta}_9 & \tilde{\delta}_{10} & \tilde{\delta}_{11} & \tilde{\delta}_{12} & \tilde{\beta}_{13} & \tilde{\delta}_{14} & \tilde{\beta}_{15} & \tilde{\delta}_{16} \\
\tilde{\varphi}_1 & \tilde{\varphi}_2 & \tilde{\varphi}_3 & \tilde{\varphi}_4 & 0 & \tilde{\varphi}_6 & 0 & \tilde{\varphi}_8 & \tilde{\varphi}_9 & \tilde{\varphi}_{10} & \tilde{\varphi}_{11} & \tilde{\varphi}_{12} & 0 & \tilde{\varphi}_{14} & 0 & \tilde{\varphi}_{16} \\
\tilde{\psi}_1 & \tilde{\psi}_2 & \tilde{\psi}_3 & \tilde{\psi}_4 & 0 & \tilde{\psi}_6 & 0 & \tilde{\psi}_8 & \tilde{\psi}_9 & \tilde{\psi}_{10} & \tilde{\psi}_{11} & \tilde{\psi}_{12} & 0 & \tilde{\psi}_{14} & 0 & \tilde{\psi}_{16} \\
\tilde{\eta}_1 & \tilde{\eta}_2 & \tilde{\eta}_3 & \tilde{\eta}_4 & \tilde{\gamma}_5 & \tilde{\eta}_6 & \tilde{\gamma}_7 & \tilde{\eta}_8 & \tilde{\eta}_9 & \tilde{\eta}_{10} & \tilde{\eta}_{11} & \tilde{\eta}_{12} & \tilde{\gamma}_{13} & \tilde{\eta}_{14} & \tilde{\gamma}_{15} & \tilde{\eta}_{16} \\
\tilde{\theta}_1 & \tilde{\theta}_2 & \tilde{\theta}_3 & \tilde{\theta}_4 & 0 & \tilde{\theta}_6 & 0 & \tilde{\theta}_8 & \tilde{\theta}_9 & \tilde{\theta}_{10} & \tilde{\theta}_{11} & \tilde{\theta}_{12} & 0 & \tilde{\theta}_{14} & 0 & \tilde{\theta}_{16} \\
\tilde{\zeta}_1 & \tilde{\zeta}_2 & \tilde{\zeta}_3 & \tilde{\zeta}_4 & \tilde{\delta}_5 & \tilde{\zeta}_6 & \tilde{\delta}_7 & \tilde{\zeta}_8 & \tilde{\zeta}_9 & \tilde{\zeta}_{10} & \tilde{\zeta}_{11} & \tilde{\zeta}_{12} & \tilde{\delta}_{13} & \tilde{\zeta}_{14} & \tilde{\delta}_{15} & \tilde{\zeta}_{16} \\
\tilde{\eta}_1 & \tilde{\eta}_2 & \tilde{\eta}_3 & \tilde{\eta}_4 & -\tilde{\gamma}_5 & \tilde{\eta}_6 & -\tilde{\gamma}_7 & \tilde{\eta}_8 & \tilde{\eta}_9 & \tilde{\eta}_{10} & \tilde{\eta}_{11} & \tilde{\eta}_{12} & -\tilde{\gamma}_{13} & \tilde{\eta}_{14} & -\tilde{\gamma}_{15} & \tilde{\eta}_{16} \\
\tilde{\mu}_1 & \tilde{\mu}_2 & \tilde{\mu}_3 & \tilde{\mu}_4 & 0 & \tilde{\mu}_6 & 0 & \tilde{\mu}_8 & \tilde{\mu}_9 & \tilde{\mu}_{10} & \tilde{\mu}_{11} & \tilde{\mu}_{12} & 0 & \tilde{\mu}_{14} & 0 & \tilde{\mu}_{16} \\
\tilde{\zeta}_1 & \tilde{\zeta}_2 & \tilde{\zeta}_3 & \tilde{\zeta}_4 & -\tilde{\delta}_5 & \tilde{\zeta}_6 & -\tilde{\delta}_7 & \tilde{\zeta}_8 & \tilde{\zeta}_9 & \tilde{\zeta}_{10} & \tilde{\zeta}_{11} & \tilde{\zeta}_{12} & -\tilde{\delta}_{13} & \tilde{\zeta}_{14} & -\tilde{\delta}_{15} & \tilde{\zeta}_{16} \\
\tilde{\nu}_1 & \tilde{\nu}_2 & \tilde{\nu}_3 & \tilde{\nu}_4 & 0 & \tilde{\nu}_6 & 0 & \tilde{\nu}_8 & \tilde{\nu}_9 & \tilde{\nu}_{10} & \tilde{\nu}_{11} & \tilde{\nu}_{12} & 0 & \tilde{\nu}_{14} & 0 & \tilde{\nu}_{16}
\end{pmatrix}
\end{equation}
Diagonalizing matrix for $\Omega_2$:
\begin{equation}
\begin{pmatrix}
\tilde{\alpha}_1 & \tilde{\alpha}_2 & \tilde{\alpha}_3 & \tilde{\alpha}_4 & \tilde{\alpha}_5                            & 0 & 0 & 0 & 0                                                                                     & \tilde{\alpha}_{10} & 0 & \tilde{\alpha}_{12} & 0 & 0 & 0 & 0 \\
\tilde{\beta}_1 & \tilde{\beta}_2 & \tilde{\beta}_3 & \tilde{\beta}_4 & \tilde{\beta}_5                                 & \tilde{\alpha}_6 & \tilde{\alpha}_7 & \tilde{\alpha}_8 & \tilde{\alpha}_9                         & \tilde{\beta}_{10} & \tilde{\alpha}_{11} & \tilde{\beta}_{12} & \tilde{\alpha}_{13} & \tilde{\alpha}_{14} & \tilde{\alpha}_{15} & \tilde{\alpha}_{16} \\
\tilde{\beta}_1 & \tilde{\beta}_2 & \tilde{\beta}_3 & \tilde{\beta}_4 & \tilde{\beta}_5                                 & -\tilde{\alpha}_6 & -\tilde{\alpha}_7 & \tilde{\alpha}_8 & \tilde{\alpha}_9                       & \tilde{\beta}_{10} & -\tilde{\alpha}_{11} & \tilde{\beta}_{12} & -\tilde{\alpha}_{13} & -\tilde{\alpha}_{14} & -\tilde{\alpha}_{15} & \tilde{\alpha}_{16} \\
\tilde{\gamma}_1 & \tilde{\gamma}_2 & \tilde{\gamma}_3 & \tilde{\gamma}_4 & \tilde{\gamma}_5                            & 0 & 0 & \tilde{\beta}_8 & \tilde{\beta}_9                                                        & \tilde{\gamma}_{10} & 0 & \tilde{\gamma}_{12} & 0 & 0 & 0 & \tilde{\beta}_{16} \\
\tilde{\beta}_1 & \tilde{\beta}_2 & \tilde{\beta}_3 & \tilde{\beta}_4 & \tilde{\beta}_5                                 & -\tilde{\alpha}_6 & \tilde{\alpha}_7 & -\tilde{\alpha}_8 & -\tilde{\alpha}_9                      & \tilde{\beta}_{10} & \tilde{\alpha}_{11} & \tilde{\beta}_{12} & \tilde{\alpha}_{13} & -\tilde{\alpha}_{14} & -\tilde{\alpha}_{15} & -\tilde{\alpha}_{16} \\
\tilde{\delta}_1 & \tilde{\delta}_2 & \tilde{\delta}_3 & \tilde{\delta}_4 & \tilde{\delta}_5                            & 0 & -\tilde{\beta}_7 & 0 & 0                                                                      & \tilde{\delta}_{10} & \tilde{\beta}_{11} & \tilde{\delta}_{12} & \tilde{\beta}_{13} & 0 & 0 & 0 \\
\tilde{\varepsilon}_1 & \tilde{\varepsilon}_2 & \tilde{\varepsilon}_3 & \tilde{\varepsilon}_4                           & \tilde{\varepsilon}_5 & -\tilde{\beta}_6 & 0 & 0 & 0                                              & \tilde{\varepsilon}_{10} & 0 & \tilde{\varepsilon}_{12} & 0 & -\tilde{\beta}_{14} & \tilde{\beta}_{15} & 0 \\
\tilde{\varphi}_1 & \tilde{\varphi}_2 & \tilde{\varphi}_3 & \tilde{\varphi}_4 & \tilde{\varphi}_5                       & -\tilde{\gamma}_6 & -\tilde{\gamma}_7 & \tilde{\gamma}_8 & \tilde{\gamma}_9                       & \tilde{\varphi}_{10} & -\tilde{\gamma}_{11} & \tilde{\varphi}_{12} & \tilde{\gamma}_{13} & -\tilde{\gamma}_{14} & -\tilde{\gamma}_{15} & \tilde{\gamma}_{16} \\
\tilde{\beta}_1 & \tilde{\beta}_2 & \tilde{\beta}_3 & \tilde{\beta}_4 & \tilde{\beta}_5                                 & \tilde{\alpha}_6 & -\tilde{\alpha}_7 & -\tilde{\alpha}_8 & -\tilde{\alpha}_9                       & \tilde{\beta}_{10} & -\tilde{\alpha}_{11} & \tilde{\beta}_{12} & -\tilde{\alpha}_{13} & \tilde{\alpha}_{14} & \tilde{\alpha}_{15} & -\tilde{\alpha}_{16} \\
\tilde{\varepsilon}_1 & \tilde{\varepsilon}_2 & \tilde{\varepsilon}_3 & \tilde{\varepsilon}_4 & \tilde{\varepsilon}_5   & \tilde{\beta}_6 & 0 & 0 & 0                                                                       & \tilde{\varepsilon}_{10} & 0 & \tilde{\varepsilon}_{12} & 0 & \tilde{\beta}_{14} & -\tilde{\beta}_{15} & 0 \\
\tilde{\delta}_1 & \tilde{\delta}_2 & \tilde{\delta}_3 & \tilde{\delta}_4 & \tilde{\delta}_5                            & 0 & \tilde{\beta}_7 & 0 & 0                                                                       & \tilde{\delta}_{10} & -\tilde{\beta}_{11} & \tilde{\delta}_{12} & -\tilde{\beta}_{13} & 0 & 0 & 0 \\
\tilde{\varphi}_1 & \tilde{\varphi}_2 & \tilde{\varphi}_3 & \tilde{\varphi}_4 & \tilde{\varphi}_5                       & \tilde{\gamma}_6 & \tilde{\gamma}_7 & \tilde{\gamma}_8 & \tilde{\gamma}_9                         & \tilde{\varphi}_{10} & \tilde{\gamma}_{11} & \tilde{\varphi}_{12} & -\tilde{\gamma}_{13} & \tilde{\gamma}_{14} & \tilde{\gamma}_{15} & \tilde{\gamma}_{16} \\
\tilde{\gamma}_1 & \tilde{\gamma}_2 & \tilde{\gamma}_3 & \tilde{\gamma}_4 & \tilde{\gamma}_5                            & 0 & 0 & -\tilde{\beta}_8 & -\tilde{\beta}_9                                                        & \tilde{\gamma}_{10} & 0 & \tilde{\gamma}_{12} & 0 & 0 & 0 & -\tilde{\beta}_{16} \\
\tilde{\varphi}_1 & \tilde{\varphi}_2 & \tilde{\varphi}_3 & \tilde{\varphi}_4 & \tilde{\varphi}_5                       & \tilde{\gamma}_6 & -\tilde{\gamma}_7 & -\tilde{\gamma}_8 & -\tilde{\gamma}_9                      & \tilde{\varphi}_{10} & -\tilde{\gamma}_{11} & \tilde{\varphi}_{12} & \tilde{\gamma}_{13} & \tilde{\gamma}_{14} & \tilde{\gamma}_{15} & -\tilde{\gamma}_{16} \\
\tilde{\varphi}_1 & \tilde{\varphi}_2 & \tilde{\varphi}_3 & \tilde{\varphi}_4 & \tilde{\varphi}_5                       & -\tilde{\gamma}_6 & \tilde{\gamma}_7 & -\tilde{\gamma}_8 & -\tilde{\gamma}_9                      & \tilde{\varphi}_{10} & \tilde{\gamma}_{11} & \tilde{\varphi}_{12} & -\tilde{\gamma}_{13} & -\tilde{\gamma}_{14} & -\tilde{\gamma}_{15} & -\tilde{\gamma}_{16} \\
\tilde{\psi}_1 & \tilde{\psi}_2 & \tilde{\psi}_3 & \tilde{\psi}_4 & \tilde{\psi}_5                                      & 0 & 0 & 0 & 0 & \tilde{\psi}_{10} & 0 & \tilde{\psi}_{12} & 0 & 0 & 0 & 0
\end{pmatrix}
\end{equation}
\endgroup
\end{widetext}
The matrices for a loop on a contour of width 1 and width 3 are identical if we choose different spin matrices --- we average over the interlayer edges connected to the edge~$\tau_i^m$. If we take the spin matrix corresponding to this vertical edge, we obtain a contour of width 1. If, instead, we include the spin matrices of the three remaining vertical edges in the layer into the product, we obtain a contour of width 3 due to the toroidal closure of the model (along the width). However, there are exceptions to this pattern, which will be discussed below.
\begin{theorem}
		In the thermodynamic limit $L\to\infty$, the Wilson loop for a contour of width 1:
    \begin{equation}
        WL^1_{i,j} = \sum_{i=1}^{16}\frac{\mu_i^k}{\lambda_1^k} \left|\left(\overrightarrow{r_i},\overline{\overrightarrow{v_1}}\right)_{S_1}\right|^2,
    \end{equation}
		For width 3:
    \begin{equation}
        WL^3_{i,j} = \sum_{i=1}^{16}\frac{\mu_i^k}{\lambda_1^k} \left|\left(\overrightarrow{r_i},\overline{\overrightarrow{v_1}}\right)_{S_2S_3S_4}\right|^2,
    \end{equation}
		where only the scalar products with eigenvectors having indices $\{1,2,3,4,6,8,9,10,11,12,14,16\}$ remain nonzero.
   
		For width 2:
    \begin{equation}
        WL^2_{i,j} = \sum_{i=1}^{16}\frac{\mu_i^k}{\lambda_1^k} \left|\left(\overrightarrow{r_i},\overline{\overrightarrow{v_1}}\right)_{S_1S_2}\right|^2,
    \end{equation}
		where only the scalar products with eigenvectors having indices $\{1,2,3,4,5,10,12\}$ and $\{8,9\}$ remain nonzero.
    \begin{equation}
    S_1 = \operatorname{diag}[1,1,1,1,1,1,1,1,-1,-1,-1,-1,-1,-1,-1,-1]
\end{equation}
\begin{equation}
    S_2 = \operatorname{diag}[1,1,1,1,-1,-1,-1,-1,1,1,1,1,-1,-1,-1,-1]
\end{equation}
\begin{equation}
    S_3 = \operatorname{diag}[1,1,-1,-1,1,1,-1,-1,1,1,-1,-1,1,1,-1,-1]
\end{equation}
\begin{equation}
    S_4 = \operatorname{diag}[1,-1,1,-1,1,-1,1,-1,1,-1,1,-1,1,-1,1,-1]
\end{equation}
\end{theorem}
\subsection{Search for perimeter-law and area-law dependences of the Wilson loop}
\begin{figure}[htbp!]
  \centering
  \begin{subfigure}[b]{0.45\textwidth}
    \centering
    \includegraphics[width=\textwidth]{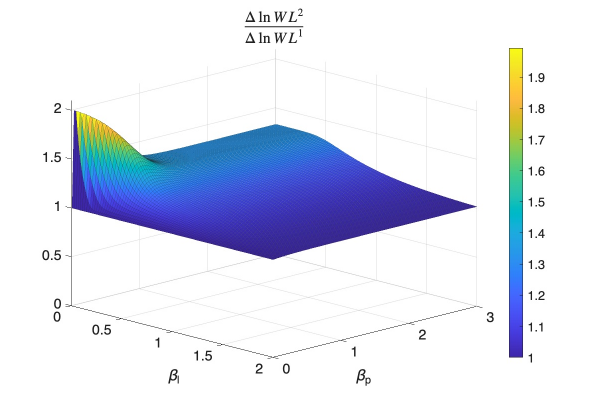}
    \caption{}
    \label{fig:surface with plaquette}
  \end{subfigure}
  \hfill
  \begin{subfigure}[b]{0.45\textwidth}
    \centering
    \includegraphics[height=4.2cm]{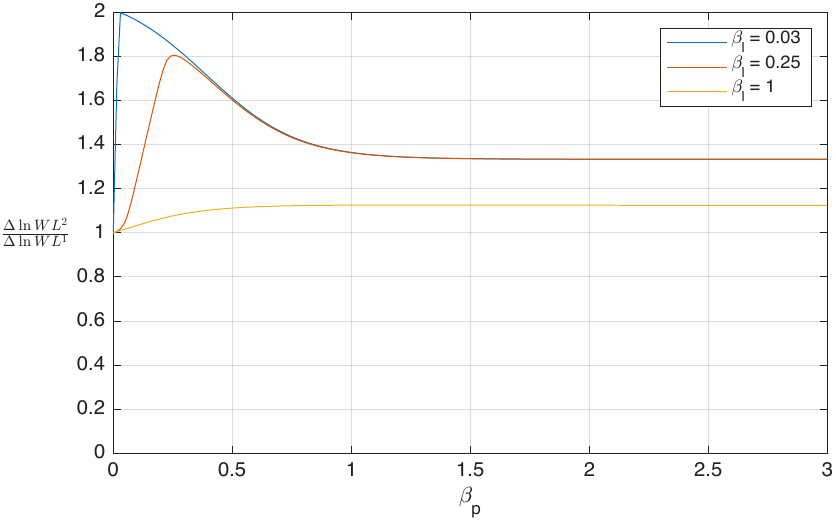}
    \caption{}
    \label{fig:plots with plaquette}
  \end{subfigure}
  \caption{Ratio of increments of the logarithms of Wilson loops of widths 2 and 1 in the form of a surface (a) with parameters $\beta_l\in[0.01,2]$ and $\beta_p\in[0,3]$ and slices (b) for $\beta_p\in[0,3]$, $\beta_l=0.03,0.25,1$.}
  \label{fig:model with plaquette}
\end{figure}

	It is worth noting that in the classical Hamiltonian with parameters $\beta_l$ for link interactions and $\beta_p$ for plaquette interactions we take into account the products in plaquettes $\tau_0^m\tau_1^m\tau_2^m\tau_3^m$, $\tau_0^{m+1}\tau_1^{m+1}\tau_2^{m+1}\tau_3^{m+1}$ with a factor of $1/2$ because they are shared by two adjacent layers. Therefore a contour covering a strip of width 3 that encloses no internal plaquettes is also a contour covering a strip of width 1 that does enclose these plaquettes at its two ends. With allowance for such interactions the exponent of the Wilson loop for a contour with three vertical edges and a contour with one vertical edge behaves identically.
\begin{figure}[htbp!]
  \centering
  \begin{subfigure}[b]{0.45\textwidth}
    \centering
    \includegraphics[height=3.4cm]{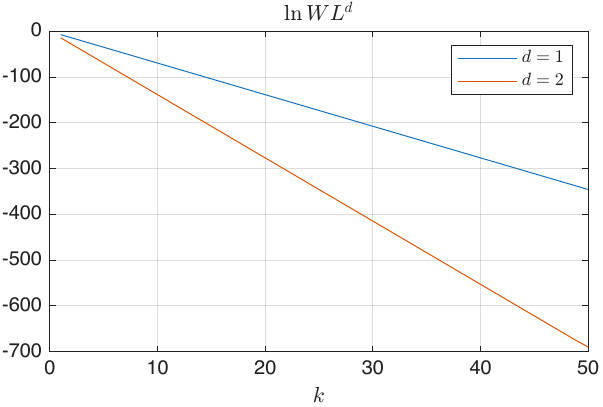}
    \caption{}
    \label{fig:surface area}
  \end{subfigure}
  \hfill 
  \begin{subfigure}[b]{0.45\textwidth}
    \centering
    \includegraphics[height=3.4cm]{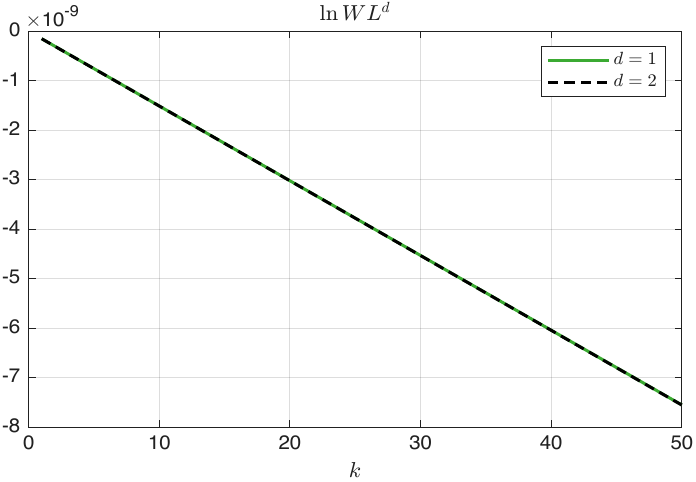}
    \caption{}
    \label{fig:plots perimeter}
  \end{subfigure}
  \caption{Plot of the Wilson loop exponent for contours of width $d$ and length $k$ at model length $L=50$, for parameters $\beta_l=0.001$, $\beta_p=0.001$ (a) and $\beta_p=4$, $\beta_l=4$ (b).}
  \label{fig:model plots}
\end{figure}\\
	On the graphs (Fig.~\ref{fig:model plots}) the dependence of the exponent is area-law or perimeter-law. Since the perimeter-law dependence does not depend on the width, the graphs in Fig.~\ref{fig:plots perimeter}(b) practically coincide. We find the ratio of the tangents of the slopes of these parameters in the space of parameters $\beta_l$, $\beta_p$ (Fig.~\ref{fig:model with plaquette}).

\newpage
\bibliography{library}  

\begin{thebibliography}{9}%
\makeatletter
\providecommand \@ifxundefined [1]{%
 \@ifx{#1\undefined}
}%
\providecommand \@ifnum [1]{%
 \ifnum #1\expandafter \@firstoftwo
 \else \expandafter \@secondoftwo
 \fi
}%
\providecommand \@ifx [1]{%
 \ifx #1\expandafter \@firstoftwo
 \else \expandafter \@secondoftwo
 \fi
}%
\providecommand \natexlab [1]{#1}%
\providecommand \enquote  [1]{``#1''}%
\providecommand \bibnamefont  [1]{#1}%
\providecommand \bibfnamefont [1]{#1}%
\providecommand \citenamefont [1]{#1}%
\providecommand \href@noop [0]{\@secondoftwo}%
\providecommand \href [0]{\begingroup \@sanitize@url \@href}%
\providecommand \@href[1]{\@@startlink{#1}\@@href}%
\providecommand \@@href[1]{\endgroup#1\@@endlink}%
\providecommand \@sanitize@url [0]{\catcode `\\12\catcode `\$12\catcode
  `\&12\catcode `\#12\catcode `\^12\catcode `\_12\catcode `\%12\relax}%
\providecommand \@@startlink[1]{}%
\providecommand \@@endlink[0]{}%
\providecommand \url  [0]{\begingroup\@sanitize@url \@url }%
\providecommand \@url [1]{\endgroup\@href {#1}{\urlprefix }}%
\providecommand \urlprefix  [0]{URL }%
\providecommand \Eprint [0]{\href }%
\providecommand \doibase [0]{https://doi.org/}%
\providecommand \selectlanguage [0]{\@gobble}%
\providecommand \bibinfo  [0]{\@secondoftwo}%
\providecommand \bibfield  [0]{\@secondoftwo}%
\providecommand \translation [1]{[#1]}%
\providecommand \BibitemOpen [0]{}%
\providecommand \bibitemStop [0]{}%
\providecommand \bibitemNoStop [0]{.\EOS\space}%
\providecommand \EOS [0]{\spacefactor3000\relax}%
\providecommand \BibitemShut  [1]{\csname bibitem#1\endcsname}%
\let\auto@bib@innerbib\@empty
%</preamble>
\bibitem [{\citenamefont {Wilson}(1974)}]{Wilson}%
  \BibitemOpen
  \bibfield  {author} {\bibinfo {author} {\bibfnamefont {K.~G.}\ \bibnamefont
  {Wilson}},\ }\bibfield  {title} {\bibinfo {title} {Confinement of quarks},\
  }\href {https://doi.org/10.1103/PhysRevD.10.2445} {\bibfield  {journal}
  {\bibinfo  {journal} {Physical Review D}\ }\textbf {\bibinfo {volume} {10}},\
  \bibinfo {pages} {2445} (\bibinfo {year} {1974})}\BibitemShut {NoStop}%
\bibitem [{\citenamefont {Wegner}(1971)}]{Wegner}%
  \BibitemOpen
  \bibfield  {author} {\bibinfo {author} {\bibfnamefont {F.~J.}\ \bibnamefont
  {Wegner}},\ }\bibfield  {title} {\bibinfo {title} {Duality in generalized
  {Ising} models and phase transitions without local order parameters},\ }\href
  {https://doi.org/10.1063/1.1665530} {\bibfield  {journal} {\bibinfo
  {journal} {Journal of Mathematical Physics}\ }\textbf {\bibinfo {volume}
  {12}},\ \bibinfo {pages} {2259} (\bibinfo {year} {1971})}\BibitemShut
  {NoStop}%
\bibitem [{\citenamefont {Balian}\ \emph {et~al.}(1975)\citenamefont {Balian},
  \citenamefont {Drouffe},\ and\ \citenamefont {Itzykson}}]{Balian}%
  \BibitemOpen
  \bibfield  {author} {\bibinfo {author} {\bibfnamefont {R.}~\bibnamefont
  {Balian}}, \bibinfo {author} {\bibfnamefont {J.-M.}\ \bibnamefont
  {Drouffe}},\ and\ \bibinfo {author} {\bibfnamefont {C.}~\bibnamefont
  {Itzykson}},\ }\bibfield  {title} {\bibinfo {title} {Gauge fields on a
  lattice. {II}. {G}auge-invariant {Ising} model},\ }\href
  {https://doi.org/10.1103/PhysRevD.11.2098} {\bibfield  {journal} {\bibinfo
  {journal} {Physical Review D}\ }\textbf {\bibinfo {volume} {11}},\ \bibinfo
  {pages} {2098} (\bibinfo {year} {1975})}\BibitemShut {NoStop}%
\bibitem [{\citenamefont {Kogut}(1979)}]{Kogut}%
  \BibitemOpen
  \bibfield  {author} {\bibinfo {author} {\bibfnamefont {J.~B.}\ \bibnamefont
  {Kogut}},\ }\bibfield  {title} {\bibinfo {title} {An introduction to lattice
  gauge theory and spin systems},\ }\href
  {https://doi.org/10.1103/RevModPhys.51.659} {\bibfield  {journal} {\bibinfo
  {journal} {Reviews of Modern Physics}\ }\textbf {\bibinfo {volume} {51}},\
  \bibinfo {pages} {659} (\bibinfo {year} {1979})}\BibitemShut {NoStop}%
\bibitem [{\citenamefont {Elitzur}(1975)}]{Elitzur}%
  \BibitemOpen
  \bibfield  {author} {\bibinfo {author} {\bibfnamefont {S.}~\bibnamefont
  {Elitzur}},\ }\bibfield  {title} {\bibinfo {title} {Impossibility of
  spontaneously breaking local symmetries},\ }\href
  {https://doi.org/10.1103/PhysRevD.12.3978} {\bibfield  {journal} {\bibinfo
  {journal} {Physical Review D}\ }\textbf {\bibinfo {volume} {12}},\ \bibinfo
  {pages} {3978} (\bibinfo {year} {1975})}\BibitemShut {NoStop}%
\bibitem [{\citenamefont {Marra}\ and\ \citenamefont
  {Miracle-Sole}(1979)}]{Marra_Miracle-Sole}%
  \BibitemOpen
  \bibfield  {author} {\bibinfo {author} {\bibfnamefont {R.}~\bibnamefont
  {Marra}}\ and\ \bibinfo {author} {\bibfnamefont {S.}~\bibnamefont
  {Miracle-Sole}},\ }\bibfield  {title} {\bibinfo {title} {On the statistical
  mechanics of the gauge invariant {Ising} model},\ }\href
  {https://doi.org/10.1007/bf01238846} {\bibfield  {journal} {\bibinfo
  {journal} {Communications in Mathematical Physics}\ }\textbf {\bibinfo
  {volume} {67}},\ \bibinfo {pages} {233} (\bibinfo {year} {1979})}\BibitemShut
  {NoStop}%
\bibitem [{\citenamefont {Turban}(2016)}]{Turban}%
  \BibitemOpen
  \bibfield  {author} {\bibinfo {author} {\bibfnamefont {L.}~\bibnamefont
  {Turban}},\ }\bibfield  {title} {\bibinfo {title} {One-dimensional {Ising}
  model with multispin interactions},\ }\href
  {https://doi.org/10.1088/1751-8113/49/35/355002} {\bibfield  {journal}
  {\bibinfo  {journal} {Journal of Physics A: Mathematical and Theoretical}\
  }\textbf {\bibinfo {volume} {49}},\ \bibinfo {pages} {355002} (\bibinfo
  {year} {2016})}\BibitemShut {NoStop}%
\bibitem [{\citenamefont {Cardano}\ \emph {et~al.}(2007)\citenamefont
  {Cardano}, \citenamefont {Witmer},\ and\ \citenamefont {Ore}}]{Cardano}%
  \BibitemOpen
  \bibfield  {author} {\bibinfo {author} {\bibfnamefont {G.}~\bibnamefont
  {Cardano}}, \bibinfo {author} {\bibfnamefont {T.~R.}\ \bibnamefont
  {Witmer}},\ and\ \bibinfo {author} {\bibfnamefont {O.}~\bibnamefont {Ore}},\
  }\href@noop {} {\emph {\bibinfo {title} {The rules of algebra: Ars Magna}}},\
  Vol.\ \bibinfo {volume} {685}\ (\bibinfo  {publisher} {Courier Corporation},\
  \bibinfo {year} {2007})\BibitemShut {NoStop}%
\bibitem [{\citenamefont {Khrapov}\ and\ \citenamefont
  {Volkov}(2026)}]{Volkov_Khrapov}%
  \BibitemOpen
  \bibfield  {author} {\bibinfo {author} {\bibfnamefont {P.}~\bibnamefont
  {Khrapov}}\ and\ \bibinfo {author} {\bibfnamefont {N.}~\bibnamefont
  {Volkov}},\ }\href {https://arxiv.org/abs/2605.17600} {\bibinfo {title}
  {Exact solution and pair correlation functions for a generalized three-chain
  {Ising} tube with multispin interactions}} (\bibinfo {year} {2026}),\ \Eprint
  {https://arxiv.org/abs/2605.17600} {arXiv:2605.17600 [cond-mat.stat-mech]}
  \BibitemShut {NoStop}%
\end{thebibliography}%
\end{document}